\documentclass[11pt,letterpaper,twoside,twocolumn]{article}

\author{David P. Chassin and Ned Djilali}

\title{Multi-scale Transactive Control In Interconnected Bulk Power Systems Under High Renewable Energy Supply and High Demand Response Scenarios: Auxiliary Report}

\usepackage[left=2cm,right=2cm,top=2cm,bottom=2cm]{geometry}

\usepackage[utf8]{inputenc}
\usepackage{amsmath}
\usepackage{amsfonts}
\usepackage{amssymb}

\usepackage{tocloft}

\usepackage{dcolumn}
\usepackage{longtable}
\newcommand{\cell}[2][c]{%
  \begin{tabular}[#1]{@{}c@{}}#2\end{tabular}}
\newcommand{\lcell}[2][c]{%
  \begin{tabular}[#1]{@{}l@{}}#2\end{tabular}}
\usepackage{pdflscape}
\usepackage{afterpage}
\usepackage{rotating}

\usepackage{amsmath}
\usepackage{amsthm}
\usepackage{amssymb}

\usepackage{ifsym,wasysym,textcomp,trfsigns,stmaryrd}

\usepackage{color,soul}
\usepackage{xspace}
\usepackage{textcase}

\usepackage{wasysym}
\usepackage{graphics}
\usepackage{graphicx}   
\usepackage{wrapfig}
\usepackage{framed}

\usepackage[all,cmtip]{xy}
\usepackage{tikz}

\usepackage{nomencl}
\makenomenclature

\usepackage{mathtools,cuted}
\usepackage{listings}
\usepackage{ulem}
\usepackage{hyperref}

\usetikzlibrary{shapes.geometric, arrows}
\usetikzlibrary{patterns}

\newcommand{\superscript}[1]{$^\mathrm{#1}$}

\newcommand{\dd}[2]{\frac{d{#1}}{d{#2}}}

\renewcommand{\H}[1]{{\mathcal H}_{#1}}

\newcommand{\reffig}[1]{Figure~\ref{fig:#1}}

\usepackage{pdfpages}

\begin{document}

\includepdf[pages=-]{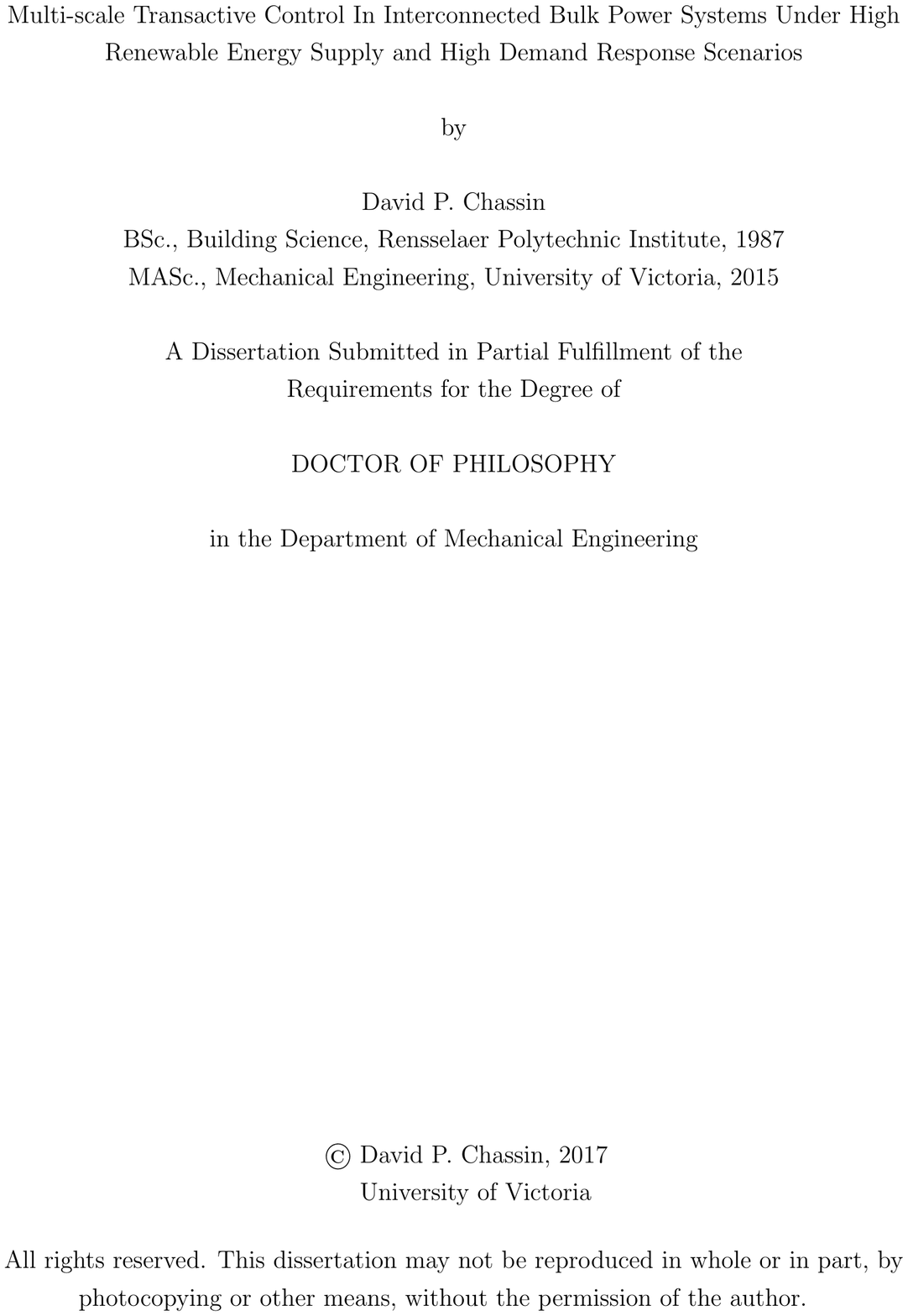}

\maketitle

\section*{Abstract}

This document is an auxiliary report to the doctoral dissertation entitled ``Multi-scale Transactive Control In Interconnected Bulk Power Systems Under High Renewable Energy Supply and High Demand Response Scenarios'' by David Chassin, presented on November 15, 2017 at the University of Victoria in British Columbia, Canada \cite{chassin2017multi}.

\section{System overview}
\label{app:response}

This section presents an overview of power system operations, the role of demand response and a history of transactive control in electric power systems. When loads act as a resource situated in electric power systems, we are motivated to ask how demand resources can be used to provide needed services in electric system planning and operations, how the individual responses of consumer devices are aggregated to enable practical application of load as a resource, and what may be the environmental and emissions impacts of employing demand response control. These questions are very complex and difficult to answer in the context of today's fast-evolving electricity infrastructure.  The changing nature of the system, the economic and regulatory context, and the loads themselves are all conditions that must be considered if we are to employ demand response to affect such as mitigating renewable intermittency, empowering consumers, and reducing the cost of system operation.

Responsibility for the reliability of electricity interconnections is shared by all the operating entities within each interconnection. In a traditional power system, these entities are vertically integrated. A committee process involving all the entities within each power pool establishes the reliability criteria utilities use for planning and operations.  Typically, the operating entities belong to larger regional coordinating councils so that they can coordinate their criteria with neighboring power pools.  Since 1965 these regional councils have been organized under what is now called the North America Electric Reliability Corporation (NERC), which establishes the standards for system reliability \cite{nerc2017standards}.

With the evolution toward market-based operations in recent decades, vertically integrated operating entities have been broken up into generation companies (GENCOs), transmission owners (TRANSCOs), load serving entities (LSEs), and energy traders that do not own assets, all of whom are collectively the market participants \cite{caramanis1982optimal}. The responsibility for ensuring the reliability of a control area is delegated to independent system operators (ISO) or regional transmission operators (RTO).  In general, market participants have the duty to provide accurate data about their assets and prices, as well as follow the dispatch orders of the ISO/RTO.  The ISO/RTO has the duty to ensure that each market participant meets the reliability rules, and determines the dispatch orders necessary for the electricity supply and demand to match according to NERC's reliability standards. This system is predicated on a successful competitive market in which private decentralized trading and investment design work to allow substantial commercial freedom for buyers, sellers and various other types of traders \cite{hogan1998primer}.

The method used to implement such a system planning and operating model uses a two-stage process referred to as the ``unbundled'' or ``two-settlement'' approach:

\begin{enumerate}

\item Unit-commitment (UC) is a days-ahead process that determines the hourly operating set points of the generation assets based on their bid energy prices and the forecast system load.

\item Economic-dispatch (ED) is an hours-ahead process that determines the real-time generation schedules and procures additional supply to ensure system reliability.

\end{enumerate}

This two-step approach can be used for both regulated and unregulated markets and the analysis method is similar for both short-term operations and long-term planning with the only caveat that ISOs must perform the system studies for deregulated markets to determine whether additional generation and transmission may be required. 

Overall the time frames for planning and operations can be separated into the following security functions \cite{chow2005electricity}:

\begin{enumerate}

\item Long term planning (2-5 years) determines needed investments in generation and transmission.

\item Resource adequacy (3-6 months) secures generation to serve expected load and sets long-term maintenance schedules.

\item Operations planning (1-2 weeks) coordinates short-term maintenance schedules and long-lead generation.

\item Day-ahead scheduling (12-24 hours) performs a security-constrained UC using energy bids.

\item Real-time commitment and dispatch (5-180 minutes) performs real-time security-based economic balancing of generation and load.

\end{enumerate}

For time intervals shorter than about five minutes, the reliability of the system is delegated entirely to the generation and loads according to reliability standards promulgated by NERC and coordinated separately by each interconnection.

\subsection{NERC Reliability Standards}

The North American power system is divided into two major interconnections, Eastern, Western and three minor interconnections: Quebec, Texas and Alaska (in approximate order of total generation capacity).  One or more reliability councils govern each interconnection.  Six reliability councils govern the Eastern Interconnection: Florida (FRCC), Midwest (MRO), Northeast (NPCC), ReliabilityFirst (RFC), Southeast (SERC) and Southwest  (SPP).  The Western interconnection is governed by WECC, Texas is governed by ERCOT, Quebec is governed by NPCC, and Alaska by ASCC (which is a affiliate member of NERC) \cite{nerc2011balancing}.

Each interconnection is operated as a single large machine. Each generator contributes in synchrony with every other generator to supply electric energy to the interconnections' customers.  The angular velocity $\omega$ of the generators at steady state determines the system frequency $f = \omega/2\pi$. However, if the power generated differs from the power consumed, the frequency will change according to the swing equation $df/dt = M \Delta P + D \Delta f$, where $M$ is the inertial constant of the interconnection, $\Delta P$ is power deviation, $D$ is a system-specific frequency correction term, and $\Delta f$ is the frequency deviation from nominal \cite{kundur1994power}. At steady state the frequency is the same throughout the entire interconnection.

Balancing Authorities (BA's), of which there are at present 131 in the United States, manage the balance of generation and load. Each BA dispatches generators to meet their individual needs, although some BA's also control loads.  The BA's are connected via high-voltage transmission lines (called tie-lines) to neighboring BA's. If one BA has too little generating capacity to support its native load, it can schedule a transfer of power from other BA's with available generating capacity within the same interconnection.  The ability to transfer power between BA's is the foundation of an interconnection's reliability. 

Because the frequency is the same across all the BA's in an interconnection, each BA must consider two inputs to the control of generation (and load, if applicable):

\begin{description}

\item[Interchange error:] net flow balance of power relative to the scheduled transfer.

\item[Frequency bias:] the obligation to provide energy to support frequency stability.

\end{description}

Each BA uses a common set of meters on the tie-lines connecting it to its neighbors to monitor and account for interchanges. Consequently all generators and loads fall strictly within the boundary of a metered region of balance control. However, only some of the generators within a particular BA are directly controlled by the BA to correct for interconnection frequency and scheduled tie-line flow deviations. 

\subsection{Balancing and Frequency Control}

Customer demand and uncontrolled generator output vary continuously within each BA resulting in unintentional deviations from scheduled tie-line flows and interconnection frequency.  Consequently, BA's must continually adjust controlled generator output by quantifying the mismatch in their interchange obligations as well as their frequency support obligations. This mismatch is estimated using the Area Control Error (ACE) in MW.  Operators in each BA monitor ACE and dispatch generation resources (and sometimes load) to keep it within limits that are generally proportional to the size of the BA.  The dispatch control is accomplished through a combination of automation, human and market (either bilateral or open) actions, and if necessary emergency actions such as automatic or manual generation or load shedding.  

ACE is to a BA what frequency is to an interconnection: when ACE is high it indicates over-generation in the BA and reflects upward pressure on the system frequency; when ACE is low it indicates under-generation in the BA and reflects downward pressure on the interconnection frequency.  However, when ACE is the same sign as the frequency error, it tends to increase ACE and when frequency error is the opposite sign it tends to decrease ACE.  This relationship is captured in the CPS-1 performance standard. Failure to maintain balance, as well as other grid problems such as congestion, equipment faults, and other rapid unilateral adjustments in generation or load will cause frequency variations that are reflected in violations of the CPS-1 standard.

BA control is maintained over a continuum of time ranging from seconds to hours and is divided accordingly into four levels of control:

\begin{description}

\item[Primary control:] The primary frequency response that occurs within seconds of a disturbance in the frequency.  It is provided by governor action on generators, which sense changes in the generators' speeds and adjusts the input to the generators' prime movers. It is also affected by certain loads such as motors whose speeds drop with frequency, by frequency protections that interrupt load at pre-defined frequency levels, and by firm load curtailment programs that ensure stability under severe disturbance scenarios. 	

Because generator loss is the most common system disturbance, the amount of Spinning Reserve in the interconnection determines the available frequency response.  It is understood that a) primary control does not restore frequency, it only arrests the frequency excursion, and b) operators must continually monitor their frequency response resources to ensure that Blackstart Units are able to control frequency and arrest excursions.	

\item[Secondary control:] Balancing services are controlled in the ``minutes'' timeframe (although some resources can respond faster; e.g., load and hydroelectric units) using the BA's' AGC system and supplemented with manual actions by the dispatchers. AGC computes the ACE and determines the most economical way for the available resources to restore balance, and sends set points to the affected generators (and loads, if any). 	

Initial reserve deployments after a disturbance can also be initiated under secondary control.  These resources maintain the minute-to-minute balance needed to restore frequency to its scheduled value following a disturbance and are usually drawn from both Spinning and Non-Spinning Reserves.

\item[Tertiary control:] These are the actions taken to set reserve resources in states that allow operators to handle current and future contingencies.  This is particularly important after a disturbance, when new reserves must deployed or restored.

\item[Time control:] Although frequency and balance control is very accurate, it is not perfect and errors due to transducer inaccuracies, problems with the SCADA hardware or software, or communications errors can lead to integral errors in frequency that must be corrected over the long term. The Time Monitor compares a clock driven by the interconnection frequency to an official reference time at NIST.  In most interconnections the Scheduled Frequency is changed when the Time Error exceeds a certain threshold. In WECC a Time Error Correction (TEC) is applied automatically through the Automatic Time Error Correction software.

\end{description}

Table~\ref{tab:services} enumerates the services provided, the time horizons over which they are relevant and the NERC standard that governs the adequacy of the service. 

\begin{table*}[!t]
	\centering
	\caption{Control continuum summary}
	~\\
	\label{tab:services}
	\begin{tabular}{|c|c|c|p{1.2in}|}
\hline
	Control & Service & Timeframe & NERC Standards 
\\ \hline
	Primary Control &Frequency Response & 10-60 Seconds & FRS, CPS1 
\\ \hline
	Secondary Control & Regulation Reserves & 1-10 Minutes & CPS1, CPS2, DCS, BAAL 
\\ \hline
	Tertiary Control & Imbalance Reserves & 10 Minutes-Hours & BAAL, DCS 
\\ \hline
	Time Control &	Time Error Correction & Hours & TEC 
\\ \hline
	\end{tabular}
\end{table*}

\subsection{ACE Calculation}

The area control error signal is called $ACE$ and is calculated by each BA using the following equation
\[
	ACE = (P_a - P_s) - 10 B(f_a-f_s) - E_m
\]
where $P_a$ is the actual net interchange (in MW), $P_s$ is the scheduled net interchange (in MW), $B$ is the BA's bias (in MW/dHz), $f_a$ is the actual frequency (in Hz), $f_s$ is the scheduled frequency (which may be offset $\pm0.02$ Hz when TEC is active), and $E_m$ is an measurement error correction factor to address inaccuracies that arise when using instantaneous flow measurements as hourly meters on tie-lines \cite{nerc2017standards}.

The actual power and frequency measurements are provided by each BA's SCADA system approximately every 4 seconds although time-synchronization of the measurements is not guaranteed \cite{kundur1994power}.

\subsection{Ancillary Service Markets}

Market designers identify system operators as the party responsible for reliability in a way that is compatible with competitive energy markets. One of the key lessons learned from California's market failure is that the definition of what constitutes an ancillary service is critical to ensuring sufficient liquidity, which influences reliability. A key aspect of the ancillary service market designs coming out of the 1990s is the recognition of cascading downward substitutability of reserve resources because faster responding resources are considered higher quality \cite{oren2001design}. Consequently price inversions (viz., slower reserves getting higher prices) are an undesirable property of ancillary service markets and can lead to perverse incentives and inappropriate bidding. Unfortunately early solutions to market design problems varied and appeared to address ISO-specific concerns rather than the broader issue of what constitutes a ``good'' reserves market design. 

Reserve markets can be viewed as a multi-part auction where resources compete to provide reserve services by submitting two bids, one for capacity and one for energy. The resource ranking is determined by the capacity price and pays all bidders the price of the last capacity resource reserved. The energy bids are only used when the reserves are called, and then all called reserves are paid the highest energy price of the reserves called. Such a market design was proven to be incentive compatible, meaning that bidders are induced to reveal their true marginal energy and capacity costs \cite{chao2002multi}.

\subsection{Current US Market Designs}

In 2012 the US Department of Energy commissioned a survey of how the seven major ISOs and RTOs operate their reserve markets \cite{ellison2012survey}.  In general these ISO/RTO's require the entities that serve loads to provide reserves in proportion to their loads. However, most of these entities do not have generation resources of their own, so they must acquire reserves through bilateral contracts or through centrally organized open markets where they exist.  To date, seven such markets have been organized in the US: CAISO, ERCOT, ISO-NE, MISO, NYISO, PJM, and SPP. 

None of these markets provide for trading of primary frequency control or time control.  Only secondary and tertiary control reserves are traded in open markets. Each of the ISO/RTOs defines which reserves can be traded in open markets slightly differently. In addition, markets often use different terms to describe similar resources and in some cases the same term can refer to different services. This has led to considerable confusion about what, when and where ancillary services can be provided in markets. Table~\ref{tab:terms} summarizes the different terms used for various secondary and tertiary control reserves. Each market has individual characteristics that distinguish it from others.  These are summarized in Table~\ref{tab:markets}.

\begin{table*}[!t]
	\centering
	\caption{Reserve market terminology in the US}
	\label{tab:terms}
	~\\
	\begin{tabular}{|l|l|l|l|}
	\hline
		Market 
	&	Secondary 
	&	Tertiary 
	&	Other reserves 
	\\ \hline
		CAISO 
	&	\lcell {
			Reg Reserve: $^{(1)}$ 
		\\	$\bullet$ Reg up 
		\\	$\bullet$ Reg down 
		}
	&	Spinning Reserve 
	&	Non-spinning Reserve 
	\\ \hline
		ERCOT 
	&	\lcell {
			Reg Svc: $^{(1)}$ 
		\\	$\bullet$ Reg up
		\\	$\bullet$ Reg down
		}
	&	Replacement Reserve Svc 
	&	\lcell {
			Non-spinning Reserve Svc
		\\	Replacement Reserve Svc
		}
	\\ \hline
		ISO-NE
	&
	& 	10-min Spinning
	&	\cell {
			10-min Non-Spinning
		\\	30-min Non-Spinning
		}
	\\ \hline
		MISO
	&	Reg Reserve
	&	\lcell {
			Contingency Reserve: $^{(1)}$ 
		\\	$\bullet$ Spinning Reserve
		}
	&	Supplemental Reserve
	\\ \hline
		NYISO
	&
	&	10-min Spin Reserve
	&	\lcell {
			10-min Non-sync Reserve
		\\	30-min Spinning Reserve
		\\	30-min Non-sync Reserve
		}
	\\ \hline
		PJM
	&
	&	\lcell {
			Contingency Reserve: $^{(1)}$ 
		\\	$\bullet$ Sync Reserve
		}
	&	\lcell {
			Quick Start Reserve
		\\	Supplemental Reserve
		}
	\\ \hline
		SPP $^{(2)}$
	&	\lcell {
			Regulation: $^{(1)}$ 
		\\	$\bullet$ Reg up
		\\	$\bullet$ Reg down
		}
	&	\lcell {
			Contingency Reserve: $^{(1)}$
		\\	$\bullet$ Spin Reserve
		}
	&	Supplemental Reserve
	\\ \hline
	\end{tabular}
	\footnotesize{
	\begin{tabular}{l}
	\\ Source: Ellison (2012).  
	\\ Notes: 
	\\ $^{(1)}$ Categories of reserve markets. 
	\\ $^{(2)}$ SPP information is based on their proposed market design.
	\end{tabular}
	}
\end{table*}

\begin{sidewaystable*}
	\caption{Characteristics of US electricity reserve market}
	\label{tab:markets}
	~\\
	\begin{tabular}{|l|c|c|c|c|c|c|c|}
	\hline
			Product Characteristics $^{(1)}$
		&	CAISO
		&	ERCOT
		&	ISO-NE
		&	MISO
		&	NYISO
		&	PJM
		&	SPP $^{(2)}$
	\\ \hline
		\lcell {
			Regulation Reserves (secondary frequency control)
		\\	$\bullet$	Governor control required?
		\\	$\bullet$	Separate up/down regulation markets
		\\	$\bullet$	AGC signal required?
		\\	$\bullet$	Maximum time to deliver called capacity
		\\	$\bullet$	Minimum duration to sustain called capacity
		\\	$\bullet$	Minimum ramp rate (MW/m)
		\\	$\bullet$	Minimum capacity offered
		}
	&	\cell {
			~
		\\	No
		\\	Yes
		\\	Yes
		\\	10-30 
		\\	$^{(3)}$
		\\	~
		\\	~
		}
	&	\cell {
			~
		\\	Yes
		\\	Yes
		\\	Yes
		\\	10
		\\	~
		\\	~
		\\	1 MW
		}
	&	\cell {
			~
		\\	No
		\\	No
		\\	Yes
		\\	5
		\\	60
		\\	1
		\\	~
		}
	&	\cell {
			~
		\\	Yes
		\\	No
		\\	Yes
		\\	5
		\\	60
		\\	~
		\\	~
		}
	&	\cell {
			~
		\\	No
		\\	No
		\\	Yes
		\\	5
		\\	~
		\\	~
		\\	~
		}
	&	\cell {
			~
		\\	No
		\\	No
		\\	Yes
		\\	5
		\\	~
		\\	~
		\\	0.1 MW
		}
	&	\cell {
			~
		\\	No
		\\	Yes
		\\	Yes
		\\	~
		\\	60
		\\	~
		\\	~
		}
	\\ \hline
		\lcell {
			Spinning Reserves (tertiary frequency control)
		\\	$\bullet$ Governor control required?
		\\	$\bullet$ Maximum time to deliver called capacity
		\\	$\bullet$ Minimum duration to sustain called capacity
		\\	$\bullet$ Minimum capacity offered
		\\	$\bullet$ Two-tiered market?
		}
	&	\cell {
			~
		\\	No
		\\	10
		\\	30
		\\	~
		\\	No
		}
	&	\cell {
			~
		\\	Yes
		\\	10
		\\	~
		\\	1 MW
		\\	No
		}
	&	\cell {
			~
		\\	No
		\\	10
		\\	60
		\\	
		\\	No
		}
	&	\cell {
			~
		\\	No
		\\	10
		\\	60
		\\	
		\\	No
		}
	&	\cell {
			~
		\\	No
		\\	10
		\\	~
		\\	~
		\\	No
		}
	&	\cell {
			~
		\\	No
		\\	10
		\\	~
		\\	~
		\\	Yes
		}
	& 	\cell {
			~
		\\	No
		\\	10
		\\	60
		\\	~
		\\	No
		}
	\\ \hline
		\lcell {
			Non-Spinning Reserves
		\\	$\bullet$ Maximum time to synchronize at called capacity
		\\	$\bullet$ Minimum duration to sustain called capacity
		\\	$\bullet$ Minimum capacity offered
		}
	&	\cell {
			~
		\\	10
		\\	30
		\\	~
		}
	&	\cell {
			~
		\\	30
		\\	~
		\\	1 MW
		}
	&	\cell {
			~
		\\	10
		\\	60
		\\	~
		}
	&	\cell {
			~
		\\	10
		\\	60
		\\	~
		}
	&	\cell {
			~
		\\	10
		\\	~
		\\	~
		}
	&	\cell {
			~
		\\	10
		\\	~
		\\	~
		}
	&	\cell {
			~
		\\	10
		\\	60
		\\	~
		}
	\\ \hline
		\lcell {
			Supplemental Reserves
		\\	$\bullet$ Synchronized and non-synchronized markets split
		\\	$\bullet$ Maximum time to synchronize at called capacity
		\\	$\bullet$ Minimum duration to system called capacity
		}
	&	\cell {
			~
		\\	N/A
		\\	N/A
		\\	N/A
		}
	&	\cell {
			~
		\\	No
		\\	Contract
		\\	Contract
		}
	&	\cell {
			~
		\\	No
		\\	30
		\\	~
		}
	&	\cell {
			~
		\\	N/A
		\\	N/A
		\\	N/A
		}
	&	\cell {
			~
		\\	Yes
		\\	30
		\\	~
		}
	&	\cell {
			~
		\\	No
		\\	30
		\\	~
		}
	&	\cell {
			~
		\\	N/A
		\\	N/A
		\\	N/A
		}
	\\ \hline
	\end{tabular}
	\footnotesize
	\\
	Source: Ellison et al. (2012). 
	\\ Notes: 
	\\ $^{(1)}$ All units are minutes unless otherwise noted. 
	\\ $^{(2)}$ SPP information is based on their proposed market design. 
	\\ $^{(3)}$ Values can be anywhere in this range.
\end{sidewaystable*}

The procurement of services in all seven ISO/RTOs is completed following a co-optimization of energy and reserve resources. The specific co-optimization methods used differ; some are integrated, some are coupled, and some are decoupled co-optimizations. PJM's co-optimization is coupled in forward markets with continuous real-time adjustments. ISO-NE supports a decoupled co-optimization in forward markets with a coupled real-time market.  ERCOT supports integrated co-optimization in day-ahead markets and coupled in real-time. MISO, NYISO and CAISO support fully integrated co-optimization of energy and reserve resource markets.

The settlement practices of the ISO/RTOs have come under considerable scrutiny from FERC.  As of 2014 only ISO-NE and NYISO had any form of pay-for-performance.  The remaining five ISOs pay only for the capacity accepted by the market. This was the motivation for FERC Order 755, which requires compensation for regulation based on actual services provided \cite{ferc2011order755}.  The impact of this order has yet to be realized in new settlement policies that conform to its intent.

Finally, FERC Order 745 addresses how demand response is compensated in wholesale energy markets (rather than reserve markets).  This order requires that dispatch of demand response be subject to a net-benefits test to determine whether it is cost-effective, and when it is dispatched that it be compensated at the market price for energy (e.g., the locational marginal price) \cite{ferc2011order745}.  The order's objective is to remove barriers to participation for demand response, but it has been criticized for not recognizing the nature of demand in organized electricity markets [14].  For example, the value of the LSE's obligation to serve is not reflected properly in the demand curve, and therefore a demand ``negawatts'' should not be priced as supply ``megawatts''. Such issues have yet to be properly addressed by regulators, market designers, market operators, load serving entities, and regulation service providers.

\subsection{Ancillary Services Using Demand Resources}

Modern bulk electric interconnections are constrained by the physical requirement that electric energy is not stored in any substantial way during system operations.  Historically, utility operations have focused on controlling generation to ``follow'' load to ensure that at every moment supply exactly matches demand and losses. To make electric utility planning and operation economical and manageable, the industry divides generation resources into three principal categories: base load, intermediate load, and peak load \cite{kundur1994power}. 
 
Base-load generation is the bottom portion of the supply stack that essentially runs uninterrupted throughout the year (except during maintenance or unplanned outages). Intermediate-load generation runs continuously but only seasonally as the diurnal load nadir rises and falls.  Peak-load generation is the supply that must be started and stopped daily to follow the diurnal load variations.  Each of these types of generation also provides regulation and reliability resources to help control frequency and respond to contingencies and emergencies in generation and transmission operations.  

For decades, load had not generally been considered part of the overall planning and operations model of electric interconnections except to the extent that its growth sets the conditions for capacity planning.  But in recent years increasing thought has been given to the role that load can play as a demand resource that a) reduces the need for new conventional generation resources, b) avoids using generation resources in inefficient ways, and c) enables the addition of generation resources that exhibit substandard performance characteristics when operating under the conventional load following paradigm \cite{rahimi2010overview}.

Today the term ``demand resource'' encompasses a wide range of products, services, and capabilities related to the control and management of load in electric systems.  Prior to the advent of ``smart grid'' technology, demand resources were primarily considered for planning purposes, such as demand-side management (DSM) programs, and very limited operational purposes such as in extremis under-frequency or under-voltage load shedding programs (UFLS/UVLS).  DSM programs focus on energy efficiency and other long-term demand management strategies to reduce load growth so that the need for significant new generation capacity investments can be reduced or deferred.  Generally these programs pay for themselves by reducing capital costs for a number of years, possibly indefinitely.  DSM programs helped the industry transition from its pre-1970s 7\% annual capacity growth to the sub-3\% growth prevalent today in modern electricity interconnections.

This section reviews some of the recent trends and the developments in demand response as a resource and how it can address a wider range of operational and economic system needs. The focus is particularly on the limitations of advanced demand response programs and the ongoing research to address these challenges.

\subsubsection{Load as a Resource}

Demand response programs have generally been divided into two major categories: incentive-based programs and price-based programs. Both categories recognize that there is an essential economic component to developing demand response capabilities in electric systems, but realize the economic benefits in very different ways.  As a general rule, incentive programs are contractual, typically bilateral arrangements between customers and system operators to provide direct load control, interruptible load, or market-based control strategies for emergency reserves and ancillary services.  In contrast, price-based programs use utility rate structures and energy prices such as time-of-use rates, critical-peak pricing, or real-time pricing to drive demand to be responsive to system conditions through economic signals as a proxy for direct control signals \cite{albadi2008summary}.

However, the ability of load to provide resources that serve system planning needs such as capacity investment deferrals or operational needs such as ancillary services is limited by (1) the intrinsic nature of the devices and equipment composing end-use loads and the constraints arising from consumer behavior and expectations, (2) our ability to control the loads in an appropriate and dependable manner, and (3) our ability to validate, verify, and meter their contributions to system planning and operation.

\subsubsection{Load Modeling}

The electric utility industry is extremely risk-averse because such a high value is placed on system reliability.  As a result, new technology is often limited by the ability of planners to simulate its effect in the planning studies used to establish system operating limits, and by the ability of operators to control these technologies in real-time.  In both cases, the challenge is not only modeling the technology itself, but also, more critically, simulating how the technology interacts with the bulk power system.  In the case of loads as resources for system planning and operation this modeling issue centers on three fundamental questions: (1) How do electric loads behave at various times of day, week, year? (2) How does end-use composition evolve over these time frames?  And (3) how does the control of loads affect these behaviors in shorter time horizons?

Load behavior is determined by both the electro-mechanical properties of the devices and equipment connected to the electric system and by the behavior of the consumers of the services they provide.  As a general rule utilities categorize loads by end-use, such as cooling, heating, refrigeration, lighting, cooking, plugs, washing, and drying in the residential sector.  In commercial buildings other end-uses such as computing, process pumping, conveying, and other services are also considered.  Daily, weekly, and seasonal load-shapes are associated with each of these end-uses to provide analysts with an empirical data set from which to estimate load under different conditions.  Load shapes have the advantage of capturing in a single data set both the electro-mechanical behavior and the consumer behavior that gives rise to the overall shape of loads \cite{pratt1990enduse}.

However, these load shapes have a serious drawback when one attempts to determine the degree to which a load changes in response to a short-term signals such as dispatch commands, real-time prices, frequency or voltage fluctuations: load shapes contain no information about the inter-temporal correlation of the loads' energy, power and ramping behavior.  Devising load models that incorporate these remains an ongoing area of research and tools such as LOADSYN \cite{epri1987loadsyn}, the WECC Composite Load Model \cite{wecc2012mvwg}, and GridLAB-D \cite{chassin2014gridlabd} partly address this problem.

Load composition models were developed to address a problem that generally does not arise when considering load behavior over hours or more.  Each load is composed of electrical subcomponents that have independently changing sub-hourly electro-mechanical characteristics.  Induction motors of different types, sizes, and control may start and stop; electronic power drives may be used; and the overall mix of static power, current, and impedance may change very quickly in response to dynamic frequency/voltage events, economic or dispatch signals, whether due to the normal internal control behavior or equipment protection subsystems.  Although the overall energy consumption on the hourly time scale may be described well using load shape data, the sub-hourly dynamics of power demand may be quite volatile and are often poorly understood. This lack of understanding can present system planners and operators with challenges for which few tools exist, as has been observed in the case of fault-induced delayed voltage recovery \cite{kosterev2008load}.

Load diversity is an emerging challenge when external control signals are applied to devices and equipment.  Under normal operating conditions, loads that cycle on and off are assumed to have high diversity, meaning that their cycles are relatively independent of bulk system conditions.  The difficulty is that diversity is a property of loads similar to entropy; it is difficult to directly observe and it can only be considered when compared to a reference state, such as the equilibrium state of a class of loads.  Conventional models of loads assume the diversity is maximal (at equilibrium). But in practice, load control strategies reduce diversity, sometimes to a significant degree.  In spite of these challenges, models that indirectly consider the entropy of certain load classes have been developed and applied to load control problems with some success \cite{fraser2000application,lu2004,aalami2010modeling}. But a comprehensive and theoretically sound model for diversity continues to elude load modelers, and this remains an open area of research.

Human behavior is a critical consideration when designing load control programs.  Utilities must consider two distinct aspects of human behavior to determine the viability and success of a load control program.  The first is customer recruiting and retention (the customer pays for electric services but may not be the same person as the consumer who uses the end-use service), and the second is real-time consumer participation. Customer expectations are set during the recruiting phase when utilities make a cost-benefit case for customers to opt into demand response programs.  (Demand response program marketing is primarily economic in nature but often includes an environmental component.)  After customer acquiescence, technology is usually deployed in the customers' facilities and consumers are presented with behavioral choices by the technology. The frequency of these choices can range from daily (e.g., postponing a load of laundry) to seasonal (e.g., resetting a thermostat). Expecting consumers to make choices more than once a day for any particular end-use is generally regarded as impractical and it is also usually ineffective to ask consumers to make choices less frequently than seasonally [28]. Mitigating consumer fatigue and providing continuous education have also been observed to be factors in ensuring that demand response programs are cost-effective and sustainable [29] [30].  Finally, utilities frequently face fairness and ``free-rider'' questions when customers sign-up for programs but provide no value to the utilities because either a) customers already exhibit the behavior sought, or b) the utility never calls on them to exhibit the desired behavior [31]. Ultimately the long-term effectiveness of demand response programs and the technologies that support them hinges on whether the customer benefits outweigh the consumer impacts. If there is any disconnect between customers/consumer short-term/long-term value/impacts, they will not remain in the program long enough for the program to pay for itself, let alone provide the anticipated system benefits to the utilities and system operators [32].

Until the advent of utility deregulation, demand response programs were the exclusive purview of utilities and regulated accordingly.  However, in regions where vertical integration has been overcome, third-party aggregation has become a viable business model for providing demand response from many smaller customers as a single homogenous capability that is easier for a utility or an ISO to interact with.  By using on-site control technology, utility service contracts, and rebate programs, aggregators can create both arbitrage and value-added opportunities from which to generate sufficient revenue.  In some cases, monopsony/monopoly conditions can emerge as a result of regulatory intervention, technology locked-in high front-end equipment costs and high back-end system integration costs [33].  A recent additional concern is that demand response aggregation is potentially subject to FERC jurisdiction to the extent that aggregators acquire and deliver resources across FERC jurisdictional boundaries or interact with ISO and RTO entities subject to FERC oversight.  Indeed, FERC has recently issued orders affecting how demand response is compensated in energy markets, which raises the question of whether and how it might intervene regarding demand response compensation in ancillary service markets [34].

Finally, load models ultimately are embodied in the simulation tools utilities use in planning studies and operational analysis, such as forecasting models and even billing systems where baseline load models are part of the contract.  New load models can take a very long time to be adopted by industry and become commercially available in planning and operations products.  For example, the Western Electricity Coordinating Council's Load Modeling Task Force began developing a new load model in 2001 but it was not adopted until the Summer 2013 cases.  In the interim a flat 20\% induction motor load was used after it became apparent that the standard load model was in part responsible for the discrepancies observed in the August 1996 outage studies [35]. Such delays can significantly reduce the impact and potential benefits of load control technology and approaches to faster load model adoption are still needed.

\subsubsection{Load Control}

Demand response as a tool for providing ancillary services relies on the ability to control aggregate loads.  The time scales over which loads can respond to dispatch signals, and the return to ``ready'' state determine the frequency and magnitude of load response as it performs desired ancillary services.  The models for load control (as opposed to load behavior) have yet to be developed.  Work to describe the frequency and amplitude response of modern loads and load controls has only recently been undertaken, and significant research remains to be done in this area [36].

A fast emerging obstacle to effective deployment of large-scale load control systems is the lack of a comprehensive theory of control for distributed systems.  Understanding how we regulate devices and systems in our environment is a prerequisite for managing those devices and systems. That understanding is captured in control theory, the body of formalisms that explain how we observe, control and verify key performance characteristics of engineered systems.  The challenge today is that although controllability and observability are well-defined for simple systems through the Kalman rank condition, and stability can be studied using methods named after such as Ziegler, Nichols, Bode, Nyquist, and others, the emergent behavior of interconnected systems has yet to be considered formally.  As a result, ad hoc models of robustness, security, and stochastic behavior have been overlayed on conventional control theory.  Physical constraints are often ignored, information flow is assumed instantaneous, and evolving network topologies are not well treated, so that only trivial problems are solved [37].  

The paradigm for larger, more complex and realistic systems continues to elude system engineers. We have yet to understand complex engineered systems well enough to design and control them, let alone exploit the new behaviors and possibilities inherent when linking previously independent systems into a more heterogeneous multi-technical complex of systems. In short, we need a new approach to controlling the large interconnected multi-technical complex that is emerging. The new approach must allow systems to adapt and evolve without individual components being redesigned, retested, and redeployed every time relevant parameters change.   Simply put, a new paradigm of control is needed for complex systems.

\subsubsection{Validation, Verification and Metering}

Using demand response as a resource for planning and operation depends on our ability to ensure that the tools we use for bulk power system control are accurate, work as designed for all conditions (both foreseeable and unforeseeable) and that we can monitor and meter the performance of these resources for both operational and business objectives. 

Model and simulation validation for very complex models such as the load models currently in use is a daunting challenge in itself.  Empirical end-use and load composition data collected by utilities degrade quickly and unpredictably as end-use technologies change, efficiency standards take hold and consumer habits evolve.  Although utilities know that consumer assessment surveys are essential to maintaining accurate load models, the cost of conducting these surveys has been historically prohibitive.  Many utilities and advocates of automated meter reading technology frequently cite improved consumer behavior data as one of the principal long-term benefits of automatic metering infrastructure. However, these benefits have yet to be demonstrated in practice, particularly as data privacy and security concerns begin to emerge [38].

Tool validation presents additional challenges, particularly when tools become multi-disciplinary and rely on hybrid numerical methods, such as agent-based solvers.  Although these analysis tools are highly realistic, they rarely have a reference model or baseline data to compare against.  As a result, confidence in these tools builds more slowly and the rate of adoption of advanced simulations is slower than has historically been true from more conventional power system analysis tools [39].

Control system verification remains an open research area for distributed control systems such as the large-scale demand response systems being designed and tested today.  Utilities historically relied on strictly hierarchical direct load control programs that used isolated and simple control structures and were easy to verify.  Systems that rely on autonomous responses or price signals are more likely to exhibit stochastic behavior that raise concerns regarding their reliability under extreme events, when they may become critical to system integrity [40].

Monitoring and metering are closely related to the question of verification and present additional challenges.  Utilities must measure how much resource is available in real-time to ensure that sufficient resources are deployed to provide the required contingency response.  So-called ``transactive control'' systems have the notable advantage that they provide resource availability data concurrently with the required resource cost data.  Finally, when events occur, utilities also need to measure which resources were actually deployed before compensating customers for their participation.  To date the designs of advanced demand response systems have largely failed to satisfactorily address either of these issues [41].

\subsection{Demand Response Aggregation Strategies}

In the previous sections, the role of ancillary services and the potential for demand response to provide such services were discussed in detail.  As intermittent generation becomes a standard element of the generation fleet, the interest in using demand response as a substitute for new controllable generation is expected to grow.  In addition, demand response has long been regarded as necessary because it reveals the elasticity of demand in a way that mitigates supply-side market power. 
 
One of the most significant obstacles to using demand response to simultaneously displace generation-centric reliability services and mitigating generator market power is the mismatch in the characteristic size, time, and uncertainty of loads relative to generators: there are relatively few easily observed generators and their characteristic response times are relatively slow compared to overall system dynamics. Loads in contrast are far smaller, far more numerous and difficult to observe, but potentially far faster acting that the overall system dynamics [42].
  
Bulk power system planning, operation and control have generally been designed to consider the characteristics of generators and treated loads as a ``noisy'' boundary condition.  Thus load control remains quite difficult to incorporate into bulk system planning and operation. In general, the approach to addressing this fundamental mismatch is to devise demand aggregation strategies that collect numerous small fast acting devices with high individual uncertainty into a few large slower acting aggregations with reduced uncertainty. While not requiring every electric customer to participate in wholesale markets, demand aggregation provides a means of increasing consumer participation in system resource allocation strategies---market-based or centrally controlled---and thus can mitigate the price of energy, capacity, or ramping services [43]. 

From an economic perspective, aggregating electricity customers can be viewed as a means of capturing consumer surplus to increase producer surplus by segregating consumers into groups with different willingness to pay.  Three general approaches are usually employed to creating consumer aggregation for either operational or economic objectives:

\begin{enumerate}

\item Economic aggregation is achieved using price discrimination methods such as different rates for different customer classes, product differentiation, and product or service bundling strategies.

\item Social aggregation is achieved using various subsidy programs, and other social group identification strategies such as environmental, green or early-adopter programs.

\item Technical aggregation creates technical structures that either directly aggregate consumers or indirectly enable economic or social aggregation.  Technical aggregation can be accomplished using service aggregators, creating technological lock-in with high barriers to entry or exit, or constructing local retail markets independent of wholesale energy, capacity, and ancillary service markets.

\end{enumerate}

This section discusses the motives, principles and practices generally employed by the electric power industry in achieving consumer (and demand response) aggregation.

\subsubsection{Economic Aggregation}

Price discrimination is an economic strategy used by sellers to capture additional consumer surplus.  Surplus is the economic benefit derived by bringing buyer and seller together to trade electricity products and services.  As long as the consumer's reservation price exceeds the producer's they are both overall better off economically if they complete the trade.  The net difference between the consumers' economic welfare with electric and their welfare without electricity is defined as the consumer surplus.  Similarly, the net economic benefit to electricity producers between producing electricity and not producing electricity is defined as the producer surplus.  It is the objective of both consumers and producers to maximize their respective surpluses, which in an efficient market results in the total surplus being maximized as well [44].

However producers recognize the some consumers have a greater willingness to pay for products and services. Consequently, producers can devise pricing strategies that divide the consumers in a way that increases their surplus but does not increase the total surplus, instead capturing some of the consumers' surplus. The most common of these is to create different rate structures for each customer sector (e.g., residential, commercial, industrial, municipal, agricultural). In theory such strategies have been shown to maximize producer surplus only when the demand curve is strictly convex toward the origin, but in practice this limitation is often ignored. Even though it may seem unfair to consumers that some pay less for the same product or service, price discrimination is regarded as a standard practice justified by the cost recovery needs of a capital intensive industry, and such practices are regularly endorsed by electric utility regulators [45].

Volume discounts are another common form of price discrimination that serve to aggregate consumer behavior. In the case of electric utilities, the most common form is the declining block rate, which recognizes that customers with a higher demand also have a more predictable peak demand than smaller customers.  The cost of operating electric power systems is driven in large measure by the cost of serving unpredictable peaks, so more predictable customers are offered discounted rates for the ``good'' behavior. In effect these customers are consuming a lower quality product: one that does not need to vary as much relative to the total and therefore costs less to produce.  An unfortunate side effect of declining block rates is that they can be a disincentive to conservation and many utilities are moving away from such rate structures.  Increasing block rates do promote conservation but this approach requires very careful analysis to predict the seasonal peak load variations. When significant numbers of customers come under such a rate, utility revenues can become much more sensitive to weather fluctuations than they already are [46].

Very likely the most well known form of price discrimination employed by utilities is product differentiation, viz., charging residential customers for energy usage and commercial/industrial customers for power capacity.  This form of customer aggregation recognizes that residential and small commercial customer behavior (e.g., individual appliance and equipment purchases) is more closely correlated with energy consumption and large commercial/industrial/agricultural customer behavior (e.g., increasing production capacity) is more closely correlated to peak power demand.  Utilities seek to have behavior and bills as strongly correlated as possible, and therefore prefer energy rates for residential and small commercial customers and power or demand ratchet rates for large commercial, industrial and agricultural customers [47].

The final form of economic customer aggregation, service bundling, is the most ubiquitous in electricity delivery.  The historically regulated nature of the utility business means that product bundling isn't thought of as a business strategy to increase revenues per se (as in the telecommunications business). Instead the capital-intensive nature of the business combined with the desire for simple billing (unlike the telecommunications industry) means that energy or power rates must include capital costs. Service bundling is considered as an appropriate net revenue volatility risk mitigation strategy, and regulated as such.  Most customers pay for only one product composed of several underlying services; e.g., energy (with capacity and reliability) or capacity (with energy and reliability).  All the underlying services that utilities provide, such as fuel price volatility hedging, capital financing, administration, and maintenance are blended into the simple price that each customer pays.  There is some discussion of utility business models that unbundle these services to achieve more economically efficient operations by revealing the customers' different demand elasticities and reservation prices for each service.  Utilities would then be able to serve customers with differentiated reliability services, for example.  Most likely the technical and regulatory obstacles to this model are why it has not gained much more than academic interest. Perhaps we can expect growing interest in areas where distribution reliability is a significant issue for some customers or technical solutions like microgrids are prevalent. But that has yet to be adequately researched at this point.

\subsubsection{Social Aggregation}

Social aggregation is based more on human behavior than economic theory and is consequently less well understood.  Utilities typically base their social customer aggregates on four types of social differentiators: income class, behavioral cross-subsidies, environmental awareness and early adopters.  
In many areas, utilities and governments provide subsidized service to low/fixed-income customers in the form of rebates, small customer rates, and special assistance programs. The reasoning is that electricity is considered an essential service in modern society that low/fixed-income customers cannot do without.  Regulators generally view low/fixed-income customer subsidies favorably, especially in communities where large numbers of such residents live. So where subsidies are not taxpayer-funded they are commonly found with investor-owner utilities and municipals. They may be less common with public utilities and cooperatives, but evidence appears to be somewhat anecdotal at this point.

Cross subsidies based on primary consumption drivers (e.g., energy vs. capacity) are an unavoidable feature of electricity services because of the strong motivation utilities have to maintain a simple blended rate for each customer class.  As a result of the design of the rates, one customer class is often effectively subsidizing another customer class, e.g., commercial customers subsidize residential or vice-versa.
Environmental/green products are a recent addition to the portfolio of social aggregations that utilities may use.  Under these programs, customers are offered the opportunity to pay a premium for their power if the utility purchases renewable energy to serve the demand. Ironically, often the marginal energy cost for resources is very near zero while the capital costs may be subsidized by tax incentives to the merchant generators.  Regardless, these programs represent a clear case of price discrimination based on the customer's socially-motivated willingness to pay more for a higher quality product.
 
Early adopter programs are another form of social aggregation that provides utilities the opportunity to test new products and services before making them available to mainstream customers.  Early adopter customers are willing to pay for a product whose quality is more uncertain in the sense that it may be better than existing products but it also might not work as intended by the utility or expected by the customer.

In all cases social aggregation strategies help utilities manage customer perceptions and expectations while continuing to meet basic business objectives such as customer satisfaction, regulatory compliance, environmental goals and research and development obligations that help the utilities adapt and adjust to changing business and technical conditions. 

\subsubsection{Technical Aggregation}

Technical customer aggregation strategies are less common in the electric utility business than might be expected for such a technology-intensive industry.  Only a few types of technical customer aggregation strategies can be readily discerned in modern utilities operations. Most notable are direct and indirect load control, service aggregators, retail markets, and technology lock-in strategies.

Direct load control is the oldest and most established mechanism used by utilities to aggregate customer demand response and make customer behavior ``work'' for the utility.  Although the quid pro quo is evident and real for direct load control programs, they are not as common as one might expect because the technical solutions tend to be rather intrusive and expensive to deploy.  Typical examples include water-heater and air-conditioning curtailment programs where load control switches are installed on customer equipment to allow the utility to directly turn off load if needed.  Customers are offered a rebate either for participation in the program (a reservation price) or for each event (a call price).  The advantage of reservation pricing is that measurement of individual responses to events is not required to properly compensate customers who participate. Unfortunately, free-rider behavior is quite prevalent under reservation pricing, particularly if the resource is rarely called.  Call pricing is more challenging because it requires measurement and verification of each individual customer response to determine compensation.  In addition, for many types of loads it can be difficult to determine what the customer would have done had the load control signal not be sent, especially if calls are frequent or continual. Hence it is more difficult to determine the appropriate compensation for each call.

Indirect load control circumvents many of these issues by avoiding direct signals to individual customers in favor of a common signal sent to all customers who have agreed to participate in the load control program.  Under such situations, reservation pricing is strongly favored, although event/call pricing remains a viable option.  The primary advantage of indirect load control strategies is that they do not require the utility to determine individual signals to send to each customer.  A single price or index signal can be sent to all customers in the program and control action in response to the signal is left to the customer, based on either economically rational expectations or contractual obligations.  The challenge for utilities is to determine the precise value of the price or index signal needed to achieve a desired level of demand response.  This problem can be addressed by the next two strategies.

Service aggregators are probably the most common technical customer aggregation strategy employed by utilities.  Service aggregators are independent third-party entities that receive a dispatch objective (e.g., an index) from their utility customer and determine how to dispatch the participants they recruited to meet the utility's objective.  The service aggregators are paid by the utility for their ability to meet the utility's objective, and part of the revenue from the utility is shared with the participants either in the form of added energy management and control equipment that can help reduce the participant's overall energy costs or in the form of an event/call rebate.

Retail capacity markets address the challenges of indirect load control using pricing signals.  The advantage of retail capacity markets is that they provide price transparency and technology neutrality.  Aside from violating some important assumptions regarding load duration and generation screening curves used in planning capacity expansion [49], the main disadvantage is that they require significant infrastructure investments both in the customer premises as well as the utility back-office systems.  Retail capacity markets can also present utilities with the same capacity expansion dilemma facing transmission system operators: In the event a distribution system constraint causes prices to rise without a corresponding rise in production cost, the utility as the market maker collects a share of the surplus which it presumably uses to finance capacity expansion [50].  However, this scarcity rent is collected only from customers on constrained feeders and not from customers on unconstrained feeders.  This goes against the fundamental tenet that the electric transmission and distribution system infrastructure itself is a public good, access to which should be priced uniformly. Ironically, any remedy to reinstate the basic nature of the public good would undermine the very demand response behavior that real-prices offer to elicit.

The final and perhaps most insidious technical strategy for customer aggregation is technology lock-in and high barriers to entry.  Although in the rational economic model the cost of the technology previously procured by a customer should be viewed as a sunk cost, customers often remain with the technology they have in spite of the existence of an objectively less costly choice going forward.  This often leads to persistent conditions where inferior technologies remain extant and causes significantly higher welfare losses than in more innovative industries [51].  Technology lock-in is often an explicit business strategy for technology and service providers to ensure that they capture a disproportionate share of the producer surplus long after more competitive technology is available.  However, in the electric power industry the process is implicit in the capital-intensive nature of the business.  That is in part the argument in favor of government-subsidized investments and technology transfer incentives in utility technology overhauls, such as for the nuclear technology in the late 1960s or renewable generation and automated metering in the late 2000s.

In the end, technical customer aggregation strategies usually support the economic, social, and business objectives of utilities and the government oversight that protects the public good portions of their operations. Technical customer aggregation is rarely an objective in itself but for various practical reasons research into technical aggregation is often divorced from these objectives.  Indeed some aggregation technologies are criticized for not recognizing these considerations and falling far short of expectations given the costs [52].

The need for utilities to aggregate customers is enduring and the methods they use vary greatly.  Utilities can employ economic and social aggregation methods to establish a robust and engaged base of customers with a greater willingness to provide demand response services. These services can be employed in electric power systems operations for energy conservation, peak load reduction and reliability services.

Although many of these aggregation methods have existed for decades, recent technological advances have enabled some of them to be revisited and enhanced.  In particular, early adopter strategies offer utilities the opportunity to test new technologies to meet regulated research program investment obligations and avoid the risk of significant capital investments, while operators and customers have to opportunity to learn how to maximize the benefits of the programs before the committing to full-scale deployment.  

Price-based strategies provide a balance of economic efficiency and risk mitigation by allowing utilities to transfer many of the costs more explicitly to customers and reducing the need to engage in price-volatility hedging on their behalf through opaque rate design processes. But regulators remain wary of price-based aggregation strategies until they can be shown to be cost-effective and fair to all customers.

\subsection{Environmental Impacts of Demand Response}

In the previous sections, the role of ancillary services, the potential for demand response to provide such services and the strategies available to aggregate demand response services were discussed in detail. We found that (1) ancillary services provide a critical capability for interconnection reliability; (2) demand response has the potential to provide such services; and (3) demand response resource aggregation is necessary to integrate such capabilities into interconnection planning and operations.
Variable (often called intermittent) generation is a growing fraction of the resource base for bulk power systems.  The variable character of certain renewable resources in particular is thought to undermine the overall reliability of the system insofar as forecasts of wind and solar generation output have greater uncertainty than more conventional fossil, nuclear or hydroelectric generation resources. As a result, the expectation is that while variable renewable generation resources do displace the energy production capacity of fossil power plants, they may not displace as much of the power or ramping capacity of those plants.  Consequently, by their variable nature, renewable resources may indicate that they do not offer as much emissions benefit as expected if one were to assess their impact simply on energy production capacity [53]. 

It seems intuitive that demand response should be able to mitigate the capacity and ramping impacts of variable generation by reducing the need to build and commit fossil generation to substitute for reserves or ramp in place of fast-changing renewable generation.  But this substitutability is constrained by (1) the nature of variable generation, the role of forecasting, and the impact of resource variability on the  emissions and economics of renewable resources; (2) the nature of load variability and how demand response is related load variability; and (3) the characteristics of end-use demand and the impact of demand response on energy consumption, peak power and ramping rates over the various time horizons that are relevant to the variable generation question.
  
Taken together, these constraints and interactions provide the basis for assessing the economic and environmental impacts of controllable load and demand response resources on various time scales. It is by virtue of the downward substitutability of reserve resources that we can assume the variability impact of renewable generation is exactly the opposite and always less than the benefit of the same controllability in demand response and we can assess the value of demand response using this inequality as a guide.

\subsubsection{Generation Variability}

On the supply side of the reliability equation we find that variability in renewable resources is the most significant contributor to uncertainty in the overall generation production scheduling process. Current renewable generation forecasting tools are based on five technologies: numerical weather prediction, ensemble forecasts, physical models, empirical modeling and benchmarking, which are combined in a 3-step process to product a forecast: (1) determine weather conditions, (2) calculate power output, and (3) scale over different time-horizons and regional conditions [54]. In general, the RMSE of renewable forecasting methods grow asymptotically as the time horizon is extended with the best models having an RMSE of less than 5\% for 1 hour forecasts to over 35\% for 3-days forecasts. There is high variability in the reported performance of different forecasting tools. Because generation resources are dispatched based on these forecasts, the principle component of unscheduled generation deviations is the error in the forecasts of renewable resources [55]. 
 
System operators schedule generation reserves primarily as a function of the amount of generation scheduled, and secondarily based on the class of generation scheduled.  Regardless of the amount or type of generation scheduled, only certain types of generation resources may serve as reserve resources, such as hydro-generation or combustion turbines.  In general, base-load thermal resources such as coal and nuclear are not usable in such a manner and it is not economical to use intermittent resources such as combined cycle plants [56].  (Renewable resources are naturally excluded from consideration because they are the source of the variability.)

The output of many types of renewable electricity generation, such as wind, wave and solar, is intermittent in nature. Output varies with environmental conditions, such as wind strength, over which the operator has no control. Including these fluctuations has the potential to affect the operation and economics of bulk systems, markets, and the output of other forms of generation. It can affect the reliability of electricity supplies and the actions needed to ensure supply always equals demand. The findings of a review of more than 200 studies on impact of intermittency provide a baseline of facts upon which we can evaluate the impact of demand response control [57]:

\begin{itemize}

\item With additional variable generation, system-level operating margins must be increased unless there is a large amount of response or controllable load. Otherwise the Loss of Load Probability (LOLP) can be expected to increase. The addition of variable generation or demand response does not change the fundamentals of how LOLP is estimated. 

\item The contribution of variable generation to reliability is measured using the capacity credit, which describes the percentage of installed capacity reduction as the share of electricity supply from variable resources. For system with less than 20\% variable generation, the capacity credit is usually 20-30\% of installed capacity, but declines as the share of electricity from variable sources increases.

\item Standby capacity is the amount by which system margin must increase in order to maintain reliability. This number only has meaning at the system level and should never be associated with any given variable resource. There is ongoing debate about whether LOLP fully captures the changes in reliability arising from variable generation because the number of small curtailment events may increase while the number of large outage events may remain unchanged or even decrease. This suggests the same may be true for using LOLP to evaluate the reliability benefits of demand response.

\item In liberalized markets, most of the electric energy is still traded primarily using medium and long-term bilateral contracts. The system operators make small residual adjustments using purchased short-term reserves. The costs of acquiring these short-term reserves are passed on to consumers. Variable generation typically adds up to 20\% to the cost of electricity supply to provide 5-10\% of the installed capacity of wind in additional supplies.  In most systems this cost is typically less than \$5/MWh.

\item In these markets, there is no single entity that is responsible for acquiring system margin and therefore the cost of additional margin required to mitigate the reliability impact of variable generation is difficult to estimate. There is also a need for a common definition of the system reliability cost of variable generation. Overall the reliability cost of variable generation is estimated at 10-20\% of its direct cost, including the cost of maintaining a high system margin but these costs do not consider the externalities of environmental and emissions impacts. For the purposes of establishing the impact equivalence between generation variability and load controllability this may be fortuitous, as it makes it unnecessary to remove these in order to establish a solid basis for cost impacts: load controls do not exhibit these same externalities, although they may have their own different ones, such as consumer behavior impacts.

\end{itemize}

Variable resources do help reduce the need to operate fossil-based power plants, and thus reduce emissions to a first order. But this benefit is not on a one-to-one basis because the need to continually adjust fossil plant output can cause second-order increases in emissions due to decreased plant efficiency. For 3 MW of wind capacity added, only 2 MW of fossil capacity is decommitted. Additional startups reduce the emissions benefits of wind by 2\%. Part-load operation reduces the emissions benefits by an additional 0.3\% in WECC [58]. In addition, at high variable generation levels, some energy may need to be spilled because there are no consumers for it under light load conditions. The effective emissions rate for wind due to these secondary effects relative to a typical interconnection fossil generation mix is about 1-2\%/MWh [59]. 

The overall emissions penalties for renewable benefits can are shown in Table~\ref{tab:emissions}. Based on the variable resource impacts inequality assumption, we should assume that demand response benefits cannot exceed these values.

\begin{table*}[!t]
	\centering
	\caption{Emission Reductions Relative to the 0\% Wind Penetration \cite{valentino2012system}}
	\label{tab:emissions}
	~\\
	\begin{tabular}{|c|c|c|c|c|c|c|c|c|}
	\hline
		Wind & CO2 & CO2 & N2O & CH4 & CO & NOx & SOx & PM
	\\ \hline
		10\% & 12\% & 12\% &  9\% & 12\% & 10\% & 13\% &  8\% & 11\%
	\\	20\% & 21\% & 21\% & 11\% & 17\% & 15\% & 22\% & 17\% & 22\%
	\\	30\% & 28\% & 28\% & 10\% & 21\% & 19\% & 29\% & 24\% & 32\%
	\\	40\% & 33\% & 33\% &  4\% & 23\% & 20\% & 34\% & 30\% & 40\%
	\\ \hline
	\end{tabular}
\end{table*}

There are a number of considerations that limit the equivalence between variable generation impacts and controllable load benefits.  In particular, the geographic dispersal of variable generation supports diversity, which is a key assumption in estimating their collective reliability impacts.  For demand response, such assumptions may not hold. In addition, certain regulatory practices such as defining gate closures (the lead time required to procure reserves) may differentially affect how well improvements in forecasting of variable generation reduce reliability impacts relative to changes in load forecasting as more load becomes responsive.

\subsubsection{Load variability}

Time-series load data is the foundation of all load analysis. The most commonly available data on load are metered balancing area, substation, feeder, premises, and end-use load data (in decreasing order of availability).  Utilities have measured balancing area to feeder-level load using SCADA systems for decades and this provides a very clear picture of the aggregated behavior of load. Most obvious in this data is the weather and diurnal sensitivity of load, which are the basis of load forecasting tools [60].  

Until recently, premises load data was only measured monthly, and depending on the rate paid by the customer it might be only energy use (so-called interval metering) or peak power (for ratchet demand rates).  However the advent of advanced metering technology has offered the possibility for significantly more detailed sub-hourly premises load data that allows analysts to examine many shorter term behaviors such as devices and equipment cycling at the sub-hourly horizon.  Although end-use metering is still very limited, it does provide additional insights that contribute important sub-hourly information to the study of load variability [61].

Recent work has identified a distinctive spectral signature for power from wind turbines [62]. The technique was successfully applied to sizing storage for variable generation mitigation [63], reducing variable generation forecast uncertainty [64], and studying load control for variable generation mitigation [65]. It particular, there appears to be an opportunity to use variability spectra to create a library of end-use load signatures that will enable the study of both load and generation variability and support the design of  demand response control programs that are better suited to mitigating variable generation. This area appears to be a potentially very fruitful topic for research with numerous opportunities, including

\begin{itemize}

\item End-use signature development for load decomposition;

\item Model identification for both duty-cycle phase and amplitude of sub-hourly load behavior;

\item Identification of human-driven behavior and demand response sensitivities; and

\item Identification of non-cyclic load variability phases and amplitudes for diurnal and seasonal behavior.

\end{itemize}

The response sensitivities based on spectral variability functions in particular appear to simplify the evaluation and analysis of variability generation and demand response impact questions.  For example, the computation of the overall emissions or cost impact of a load shift of  hours can be estimated by the convolution
\[
	U(t) = \int_{-\infty}^{+\infty} \upsilon(\tau) L(t-\tau)d\tau = (\upsilon*L)(t)
\]
where $\upsilon(t)$ is the cost of emissions at the time $t$ and $L(t)$ is the load.  While in time domain this can be difficult to compute, in frequency domain it is relatively simple:
\[
	\hat U(s) = \hat \upsilon(s) \cdot \hat L(s)
\]
where $\hat U(s)$, $\hat \upsilon(s)$, and $\hat L(s)$ are the Fourier transforms of $U(t)$, $\upsilon(t)$, $L(t)$  respectively. Given a library of both generation variability and load control signatures in frequency domain, the optimal demand response design problem may be relatively simple to evaluate.

\subsubsection{Characteristics of load}

Loads and load control exhibit a peculiar characteristic that is often not considered in benefits analysis.  The relationship between energy, load, and ramping is actually quite robust. Most demand response programs can exclusively affect either power demand in the short term or energy consumption in the long term.  In every other respect energy, power, and ramping are strictly related to each other as
\[
	\dd{}{t} Energy(t) = Load(t) = \int Ramp(t)~dt 
\]
and this relationship is not affected by conventional demand response control strategies.  For example, a DSM program may reduce energy consumption in the long term, but the power and ramping impact are strictly a function of how the demand response program affects energy use. Similarly, an air-conditioning load curtailment program to cut peak may reduce power during peak hours, but the natural tendency of thermostatic devices to make up for short-term deficits over the long run means that long term energy use may be relatively unchanged. The characteristic time of a demand response control strategy and how the systems it controls respond are essential to understanding how well demand response will mitigate variable generation resources and the degree to which the demand response impact inequality will apply.

The argument can be made that resources with greater ramping capabilities should be considered higher quality reserve resources.  In ancillary services markets, this characteristic places a premium on faster resources with downward substitutability.  For this reason, demand response resource that control the power of loads are at least as valuable as generation resources with the same net power response and often more valuable because of their greater ramping response (strong downward substitutability).  In fact, it seems that the principal (and perhaps the only) limiting factor on the ramping rate of demand response resources is the telecommunications latency of the control signals.  The real-time market in the Olympic study had a typical delay of a few seconds in response to the market clearing event, but the market itself cycled only once every five minutes [66].

\subsection{Summary of Impacts}
The impacts of generation variability hence load controllability may be summarized as follows:

\begin{itemize}

\item Long term load forecasts have lower relative RMSE than long term than variable generation forecasts. Thus load can be expected to outperform the generation it mitigates, all other things being equal.

\item Load control can be scheduled with greater reliability than variable generation and thus can be expected to outperform the generation it mitigates, all other things being equal.

\item The LOLP impacts of variable generation are mitigated by load control in part by moving all controllable load out of the load impacts by outages.

\item The capacity credit for controllable load can be expected to be comparable to the capacity credit for variable generation, if not better, because for every 1 MW of load that is controllable, 1 MW of generation reserve can be decommitted.

\item The standby capacity reduction associated with controllable load should in principle be 100\% of the active load under control.  

\item When controllable load is dispatched under liberalized markets, consumers become the providers of resources.  This tends to divert revenue from generators to savings by consumers.  Based on the cost of variability on the supply side, this can be expected to be about 10-20\% of the direct cost of electricity, and mitigates the need to provide 5-10\% additional installed capacity.

\item The secondary emissions benefits for avoiding startup and part-load fossil generation are expected to be 10-20\% for modest levels of variable generation (i.e., <20\%) but may be significantly lower for some bulk systems, depending on conditions.  

\item The geographic sensitivity of load is different and very likely less than it is for variable generation.  Loads tend to be more uniform and better diversified than variable generation.

\end{itemize}

\subsection{Recent Trends}

The trend toward a more integrated and interconnected complex energy system is inexorable.  Progress on the 21\superscript{st} century's infrastructure of complex interlocking energy resource, transformation, information, service, social, and economic networks is challenging our current understanding of these systems and our ability to design and control them. 

Significant challenges and research opportunities remain in load modeling and simulation, understanding the impact of consumer behavior on demand response, the foundational theory for controlling widely dispersed demand response resources, and the verification, validation, monitoring, and metering of demand response systems in utility operations.

Overall, it is clear that we are entering a period of increased electric utility receptiveness and growing innovation in the methods and strategies for turning a largely passive customer base into an active part of electric system operation.  Technical innovation based on sound economic and social objectives as well as robust engineering design will be instrumental in bringing about this transformation.

The impact of controllable load on system operation can be deduced from studies on the impact of variable generation.  The studies to date suggest that variable generation has both costs and benefits, and that the benefits outweigh the costs for reasonable mixes of variable generation relative to conventional resources.  

Many of the adverse impacts of variable generation are positive impacts for controllable load in the sense that the magnitude of the cost or impact as a function of generator variability is a cap on the magnitude of the benefit of load as a function of load controllability.

Controllable load exhibits the further advantage of high downward substitutability and thus can be significantly favored under liberalized ancillary service markets.  This feature of controllable load suggests that well-designed ancillary service markets along with market-based load control strategies could be a very powerful combination.  

Significant further research on how to structure such energy and ancillary service markets, design load control strategies, and model the systems in which they operate is required to further elucidate the benefits of this approach. Ultimately our ability to plan and operate bulk power systems that utilize such resources will depend on our ability to understand both the system as a whole as well as the details of the economic, electromechanical and human components which comprise it.

In March 2017, a study of demand response resources in California's three investor-owned utilities was published \cite{alstone2017california}. The authors evaluated the potential size and cost of future demand response resources. They addressed two fundamental questions: (1) What demand response services can meet California's future grid needs? And (2) what is the size and cost of the expected resource base for these demand response services?

Recognizing that demand response operates across a range of timescales, they proposed a new framework for analysis studies such as theirs. They developed a supply curve modeling framework to express the availability of system-level grid services from distributed resources based on automated metering data. They created a taxonomy and a framework that groups these services into four core categories to facilitate cost/benefit analysis:

\begin{description}

\item[Shape:] Demand response that reshapes customer load profiles through price response or on behavioral campaigns with advance notice of days to months.

\item[Shift:] Demand response that encourages the movement of energy consumption from times of high demand to times of day when there is a surplus of renewable generation. Shift could smooth net load ramps associated with daily patterns of solar energy generation.

\item[Shed:] Loads that can be curtailed to provide peak capacity and support the system in emergency or contingency events--at the statewide level, in local areas of high load, and on the distribution system, with a range in dispatch advance notice times.

\item [Shimmy:] Loads that dynamically adjust demand on the system to alleviate short-run ramps and disturbances at time scales ranging from seconds up to an hour.

\end{description}

The study confirmed that the focus on load shedding to reduce peaks should be redirected to focus more on local and distribution system needs, supporting control technology and business relationships that combine targeted fast shed with shift. This will likely require integration between policy at the CPUC and CAISO to ensure that market designs are matched with the most cost-effective pathways for demand response services. Continued work on how integrated energy efficiency, behind-the-meter storage, and demand response can lead to value across a range of categories---integrated demand-side management.

Further development may focus on efforts to integrate benefits at the system scale, on the distribution system, and at the site level using distributed resource planning. The authors did not undertake a detailed study of site-level electric bill impact or explicit distribution system service modeling dynamics. However, they did include a set of first-order estimates for the scale of benefits in these areas that are likely achievable when DR technology provides multi-scale service. Given the co-benefits for site-level service, the reported an increase of about 4 GW of additional demand response shedding capacity compared to a model run without co-benefits.

\section{Transactive Control}
\label{app:transactive}

This section presents a more or less chronological history of the development of what is now called ``transactive control'', beginning with the original conception of a homeostatic grid proposed by Schweppe et al. \cite{schweppe1980homeostatic} in which a system of autonomous devices provides support to the grid by regulating their demand using ambient signals like system frequency and local voltage. This significance of load resources was expanded on by Ihara \cite{ihara1981} when he described the physical basis of cold-load pickup behavior and showed how the intrinsic integral error feedback behavior of thermostats limits the extent to which thermostatic control of loads for demand response can be used as a resource akin to energy storage. The paper also shows how important consideration of the diversity, showing that even a first-order thermal model of individual homes yields a ring-down behavior after load service is returned. The paper also distinguishes between load outages (or curtailments) that do not cause discomfort (i.e., indoor temperatures stay within the thermostat deadband) and those that cause discomfort (i.e., stray outside the deadband).  Finally, the paper also recognizes the difficulty of connecting individual-based controls to aggregate behavior, which is a problem that continues to be explored to this day in a wide variety of ways, e.g., in \cite{lu2004,perfumo2011model,vrettos2014demand}. 

Transactive control adopts fundamental ideas about using pricing as a mechanism to optimally allocate energy resources in bulk power systems and attempts to apply them to distribution systems, with the objective of facilitating the participation of retail loads to the same extent that generators participate at the wholesale level.  The original wholesale markets developed in Chile and New Zealand in the 1980's were prototypical examples of energy-only markets. Energy-only markets have since been augmented significantly in systems such as PJM, CAISO, and others, where capacity and ancillary services markets now exist as well.  However, transactive control researchers have yet to described how markets for non-energy resources can be developed and operated at the distribution level.

The problem of pricing energy for electricity networks when loop flows are present was address by Hogan in 1992 \cite{hogan1992contract}, but has yet to be completely addressed in transactive systems.  In Hogan's paper the concept of contract network was introduced to identify the contract paths that connect short-term efficient prices for electric energy with long-term power capacity prices on tranmission systems.  Here the sense of a contract network is somewhat different from that offered by Smith \cite{smith1980contract}, where a contract ``net'' described the mechanism by which agents negotiated for prioritized access to a constrained computation resource. But the differences arise from the particulars of the resources and the system being optimized.  The underlying concept of using a bottom-up approach that respects the physical limitations of the system is basically the same. In the case of Hogan's proposal, payments to the holder of a long-term capacity contract are just the amount that make them indifferent to power delivery or compensation through a settlement.  As a result, the discovery of the price through a real-time pricing mechanisms produces the same result as a secondary market would but avoids the necessity of implementing an explicit capacity trading mechanism.

Huberman and Clearwater at Xerox PARC \cite{huberman1995} extended the contract net concept to commercial buildings. In a field demonstration they showed that the mechanism found an equitable solution for satisfying thermal comfort problems when energy supply resources were constrained. The computational mechanism employed a blind double-auction in which agents bid on behalf of energy suppliers and consumers at a given price and avoided hidden costs such as excessive actuation. 

These ideas were adopted by the Pacific Northwest National Laboratory in its demonstration for the US Department of Energy on the Olympic Peninsula \cite{hammerstrom2007pacific1,hammerstrom2007pacific2}.  In this field demonstration, over a hundred residences, two office buildings, industrial loads and distributed generation resources were provided bidding agents to interact with a retail double-auction on their behalf.  Devices were equipped with underfrequency load shedding controls, thermostatic controls, and other end-use controls that interfaced with bid/response controllers. The results showed significant increases in demand response as well as improved coordination of distributed resources. Among the benefits observed, the most significant were a 60\% reduction in short-term peak load and a 15\% reduction in long-term peak load.  

Huang et al \cite{huang2010analytics} describes the design of the Pacific Northwest SmartGrid Demonstration (PNWSGD) project, also funded by the US Department of Energy under ARRA in 2009.  This project involved 60,000 customers from twelve utilities in five states in the northwest region of the US. The project implemented an end-to-end system from generation to consumption, built around an infrastructure of newly deployed smart meters. The objective was to demonstrate how transactive control can coordinate distributed generation and demand response, and test a hierarchical but decentralized control system where each node of the power grid used local signals of demand and price to match supply with demand at varying frequencies of up to every five minutes or less. 

Melton et al. \cite{melton2012transactive} summarized the transactive control research under the PNWSGD Project. In the context of the project transactive control may be thought of as extending the notion of locational marginal pricing throughout the power system from generation to end-use. The transactional nature of the technique, however, introduces a new element in the use of a pair of signals to implement an equivalent to market closing distributed in space and time. This was achieved using a negotiation process, but it was shown to not always converge. The PNWSGD project was very ambitious and was largely successful although it did not achieve all its stated objectives.  Some utilities were not fully successful in deploying all their technologies, and some deployments did not provide the expected financial or operational returns \cite{hammerstrom2015pacific}. The final report does not make any determination regarding the causes of problems that were identified, and made no attempt to diagnose them.

Modeling and simulations of these systems remain a crucial requirement for transactive program development, and require tools that can model simultaneously the markets, the power system, and the end-use loads. Fuller et al. \cite{fuller2011analysis} examined demand response and dynamic pricing programs, which have played increasing roles in the modern smart-grid environment. The authors argue that price-driven response programs are only a relatively recent development. While active markets may allow customers to respond to fluctuations in wholesale electrical costs, they may not allow the utility to control demand as precisely as more conventionally deployed direct load control systems. Transactive markets using distributed controllers and a centralized auction can create an interactive system that may limit demand during congestion events, but otherwise does not subject consumers to utility control during normal operating periods. The advent of computing and communication resources has created the opportunity to deploy transactive demand response programs at the residential level, where the combination of automated bidding and response strategies, consumer education programs, and new demand response programs give the utility the ability to reduce demand and congestion in a more controlled manner.  

The transactive system protocol and dynamic control mechanisms require modelers to capture load variation, stochastic signal losses, consumer fatigue and limits on control signals that arise out of physical constraints. Jin et al. \cite{jin2012simulation} showed that the control mechanism can perform adequately in adjusting the aggregate supply-demand mismatch, and is robust to steady transactive signal losses. They developed a large-scale network simulation model for evaluating such a hierarchical transactive control system as part of their work on the Pacific PNWSG demonstration. In this simulation the transactive control system communicates local supply conditions using incentive signals and load adjustment responses using feedback signals in a distributed fashion in order to match the consumer-desired load to the utility-desired supply scenario.

Transactive technologies include such things as ``micro-tagging'' of resources, more use of distributed intelligence, and dynamic transaction routing and approval. Ipakchi et al. \cite{ipakchi2011demand} presents a broad concept for the Smart Grid of the future that requires coordinated management of large numbers of distributed and intermittent resources, while maintaining high degrees of grid reliability, cost-effectiveness and efficiency. This concepts involves a high-degree of information exchange between operating entities, devices, and users to facilitate scheduling, dispatching and control of intermittend generation, energy storage and demand response resources using a variety of energy and ancillary services markets. The authors call for new methods that provide real-time end-to-end management of the system. They argue that deregulation of wholesale electricity markets provides a framework in which to consider this more easily. But they also argue that recent technical advances facilitate extending this framework to a ``transactive framework'' that supports scheduling, dispatching and control using competitive markets.  

Additional resources can be made available in such systems by using conventional commercial building automation systems. Transactive control has been shown to apply to any thermostatically controlled device with one- or two-way communication, and shows qualitatively the value of the market-based controls. Katipamula \cite{katipamula2012smart} makes the case that as electricity demand continues to grow, traditional approaches to meet the demand would require significant additional demand response resources, and that transactive control of residential resources is not sufficient. Part of the solution must come from ``smart'' commercial buildings as well. As with most residential buildings, many commercial buildings lack the necessary infrastructure to participate in transactive systems. Building on the work of Huberman \cite{huberman1995} and others, the authors describe market-based transactive controls that can be implemented in an existing building automation system (BAS) with little or no additional capital expenditure and show how one can make commercial buildings more demand responsive. 

Pratt \cite{pratt2012transactive} presents the key motivations and design considerations behind the development of a real-time pricing tariff approved by the Public Utility Commission of Ohio (PUCO) for AEP's gridSMART demonstration project \cite{widergren2014aep}. The design of a revenue-neutral real-time price rate that reflects locational marginal prices from the wholesale market addressed the key desire to combine both wholesale and retail congestion costs at the distribution feeder level. This simultaneously and seamlessly managed peak loads at both the feeder and system levels. The tariff included an attempt to address key issues related to equity and providing credits and incentives for congested periods by providing a congestion rebate. The authors estimated the expected impacts on customer bills, which were considered by the PUCO when it approved the real-time price tariff.

In addition to pricing, consumers can be rewarded for tolerance to service delays and reductions, especially when engaging micro-grids and electric vehicle-to-grid services. Scaglione et al \cite{scaglione2012queuing} describes Demand Side Management (DSM) and Demand-Response (DR) programs that are aimed at revealing the intrinsic elasticity of consumer electricity demand and make it responsive to the near-term cost of supplying generation. The authors argue that DSM and DR are indispensable for balancing the market power of generators and reducing the need for reserves. But they also recognize that the debate on the right approach to integrating DSM and DR in system planning and operation is not yet closed, with transactive mechanisms being a leading contender among approaches that can aggregate smart loads, properly account for inconvenience costs, and modulate the total demand time optimally while converging to the energy dispatch. 

Widergren et al. \cite{widergren2014residential} argue that the most exciting aspect of the smart grid vision is the full participation of end-use resources with all forms of generation and energy storage in the reliable and efficient operation of an electric power system. Engaging all of these resources in a collaborative manner that respects the objectives of each resource, responds to global and local constraints, and scales to the large number of devices and systems participating is an unsolved problem. 
The American Electric Power Northeast Columbus gridSMART RTPda project demonstrated distributed decision-making system approaches. As a multi-feeder extension of the Olympic project, the Columbus project demonstrated a scale-up of residential demand response that uses the bidding transactions of supply and end-use air conditioning resources communicating with a real-time, five minute market that balanced the various needs of the participants on a distribution feeder. Running as a summer peak-load system in PJM ISO territory, the project provided valuable additional field data that complemented the Olympic data set.

The problems (and solutions) are not all technical.  Sahin and Shereck \cite{sahin2014renewable} classified the costs and benefits of renewables for all market participants using the Transactive Energy Framework proposed by the GridWise Architecture Council in 2013 \cite{melton2013gridwise}. They raise some of the concerns that restructuring the system pose.  As renewable generators start to replace conventional resources and more and more customers begin to produce their own energy locally, there is a growing concern over market access parity. As more consumers become ``prosumers'' and the utility is left with a diminished role and must maintain the grid with declining revenues. Those who cannot afford their own renewable sources pay for critical infrastructure costs with higher rates. They conclude that the current market mechanisms cannot properly distribute the costs and ensure grid reliability. 

Many important questions remain outstanding regarding the best approach to improving the grid. Bowes and Beehler \cite{bowes2015defining} make the case that the concept of an integrated grid is the natural next step in the evolution of bulk power systems. New fast and ubiquitous digital communications technology and evolving regulatory policies enable both an integrated power grid and a transactive energy system. The integrated grid uses legacy electric systems as the platform for incorporating all the new distributed and renewable resources. But they ask if consensus is reached on whether and how the legacy system allows planners and operators to adopt new technologies that are so physically and financially inter-dependent at a scale and level of performance otherwise considered impossible. Is it a safer, more  reliable and more affordable system? Have we established the value proposition for a highly integrated system that makes transactive energy possible and eventually desirable? Can we leverage all the available resources to improve the utilization of all the legacy grid assets?

Syed et al. \cite{syed2015ancillary} examined the role of demand side management in providing ancillary services to the network. However, the use of demand resources for ancillary services has traditionally been limited because of a lack of field demonstrations that test whether we can rigorously quantify their ability to support grid reliability requirements. The provision of fast-acting frequency control from demand-side resource was simulated using Kok's PowerMatcher \cite{kok2005powermatcher} in combination with real-time power control hardware. In a parallel study \cite{syed2015demand} they argue against assuming that all the flexible devices within the network must be managed and controlled under one demand-side management scheme. They explore what happens when two independent demand side management schemes control a portfolio of flexible devices. They use their findings to propose a methodology to analyze the performance of non-homogeneous control schemes using their real-time power hardware-in-the-loop co-simulation platform, and recommend this type of co-simulation as the basis for investigations of ancillary service benefits.

Sandoval and Grijalva \cite{sandoval2015future} proposed a platform to coordinate distributed energy resources as part of a building model framework to help integrate renewable resources. The platform is based on a decentralized approach the employs the concept of electricity prosumers, economic agents that both produce and consume grid products and services. An important aspect of their approach is a layered architecture reminiscent of the network stack in the Internet: a  physical layer, a local control layer, a cyber layer, a system control layer for economic dispatch and real-time control, a market transactive layer for integration and a business layer for costs and revenues. The platform enables coordination between spatially and temporally distant systems with heterogeneous resources. The coordination is maintained dynamically while accommodating both end-user and system constraints. Their results show that using such a coordination platform supports higher amounts of renewable energy while it reduces carbon emissions and operational costs.

Such highly interactive coordination requires more reliable and higher throughput communication infrastructure. In considering the impact of the Internet on power systems, Collier \cite{collier2015emerging} quotes Bob Metcalfe, the inventor of the ethernet and well-known technology visionary: 
\begin{quote}\textit{Over the past 63 years, we met world needs for cheap and clean information by building the Internet. Over the next 63 years, we will meet world needs for cheap and clean energy by building the Enernet.}
\end{quote}
The analogy to the impact of the Internet on information technolocay has created high expectations for energy networks. Revolutionary advances in electronics, telecommunications and computing technologies, and devices and applications are expected to transform how we design and operate bulk electric power systems. What started as an information network connecting people is now a system of systems connecting devices as well---an Internet of Things. According the Collier, the U.S. electric utility model is arguably becoming non-viable and may be supplanted by many smaller interconnected networks of increasingly autonomous systems with literally millions of distributed generation, storage, and energy management nodes. This new more distributed but still highly interconnected grid of smart interacting devices is growing more ubiquitous, powerful, economical, and secure. Collier argues that the Internet of Things is the \textit{sine qua non} platform for the future smart grid or, using Metcalfe's phrasing, the ``control plane for the smart grid''. 

In a lecture to the American Control Conference in 2016, Jakob Stoustrup \cite{stoustrup2016closing} described what he termed an augmented ``Transactive Control and Coordination'' framework that builds on Collier's model as well as the previously demonstrated transactive control paradigm.  This operating model was tested in the large-scale demonstration in the Pacific Northwest region of the United States \cite{hammerstrom2015pacific}.  The results exhibited many of the characteristics predicted from models and simulations made prior to the start of operations. But he also argues that such demonstrations have only served to illustrate the need for a more rigorous understanding of closed-loop behavior and more systematic approaches for choosing appropriate control parameters in such a framework. System modelers and designers still lack the theoretical underpinnings needed to fully understand the impact of closing the loop around market-like mechanisms using many local conventional autonomous closed-loop systems built on a large number of aggregated controllable devices.

Behboodi et al. \cite{behboodi2016integration} discuss the integration of plug-in electric vehicles in smart grids from different perspectives. In order to achieve a grid-friendly charging load profile, a strategy is proposed based on the transactive control paradigm. This charging strategy enables electric vehicle owners to participate in real-time pricing electricity markets to reduce their charging costs. Then, the impact of large-scale adoption of electric vehicles on electricity generation and inter-area flow schedules is discussed. In order to quantify potential changes, an interconnection-scale optimal scheduling problem is used to determine hourly tieline flows. Given price sensitive loads, the objective function of the scheduler maximizes the total social welfare. Finally, fast-acting demand response for frequency regulation is used to reduce the need for generation ramping. This supports high penetration levels of intermittent renewable resources by maintaining the short-term balance of energy supply and demand. 


Galvan et al. \cite{galvan2016transactive} developed the concept of ``transactive energy'' as an instance of transactive control for efficient electric vehicle grid integration and management. The goal is to minimize the charging cost of EVs and mitigate the adverse effects of intermittent generation resources. The study recognized also that electric vehicle charging  can result in undesirable behaviors such as transformer overloads and aggravated evening demand peaks. Using the transactive energy paradigm, a distribution system operator generates ``distribution locational marginal prices'' at constraint nodes that are sent to each customers house. Consumer needs and resource availability are updated allowing the operator to recalculate prices in response to changing patterns of supply and demand. 

One important benefit of transactive control is an improvement in the self-healing and self-managing aspects of system operations. Patterson and Geary \cite{patterson2016realtime} explore these concepts for distributed networks; e.g., renewable energy nano-, micro- and macro grids. Working up from the building level through utility scale systems, and including aggregators and system operators, they describe how power sources, loads and storage of all types that are interconnected can facilitate real-time transactional power management of power flows in the network grid. They propose a variety of practical aspects of real-time transactive energy management, including software that uses modern communications technology to host the development of a new energy marketplace. They suggest that we require a platform to integrate system planning and operations, including fault detection/location, automated feeder and line switching, and automated Volt/VAR control. If such a platform enables market-based transactions by communicating in real-time with equipment, using any communication network, then this effectively turns every transactive device in the system into a revenue-grade point-of-sale terminal.

This concept necessarily should include the growing number of electric vehicles. Behboodi et al. \cite{behboodi2016electric} argue that smart grids can help Plug-in Electric Vehicles (PEV) manage their load in a grid-friendly way. They consider the case of PEVs participating in a retail double auction electricity regulation market using transactive control. Price-responsive charging of PEVs are modeled under real-time retail price signals. Then PEVs can defer charging or even discharge when the retail prices are high enough, e.g., when feeder capacity constraints are active. For the most advanced charging strategies, as the price rises, demand from PEVs drops, and if the constraint causes further price increases, the PEVs can begin to supply energy. The results show that when rooftop solar energy is available, transactive bid-response vehicle charging strategies significantly enhance short-term electricity demand elasticity and can reduce consumer energy costs by more than 75\% in comparison to the unresponsive charge case.

But these strategies do not come without challenges for the utilities.  Localized congestion problems are expected, particularly as rooftop solar PV and fast EV chargers become more prevalent. Hu et al. \cite{hu2016application} developed a network-constrained transactive control method to integrate distributed energy resources (DER's) into a power distribution system with the purpose of optimizing the operational cost of DER's and power losses of the distribution network, as well as preventing grid problems including power transformer congestion, and voltage violations. A price coordinator is introduced to facilitate the interaction between the distribution system operator (DSO) and aggregators in the smart grid. Electric vehicles are used to illustrate the proposed network-constrained transactive control method, and the problem is solved using mathematical models that describe its operation. Using simulations, they show the effectiveness of the proposed method, and particularly how it guarantees optimality.

Rahimi and Albuyeh \cite{rahimi2016applying} consider lessons from transmission open-access with the increasing availability of distributed energy resources and smart end-use devices finding their place as transactive agents in the electric system. As the transactive paradigm at the retail/distribution level grows to accommodate new market participants, they see a number of parallels between the deregulation of the wholesale transmission sector in the mid-1990's and the opening up of the distribution grid to accommodate these new participants. This includes new notions such a Distribution Locational Marginal Prices (DLMP), which function much like wholesale LMPs, particularly in distribution systems where flow reversal and meshed flow are likely to occur. 

Amin et al. \cite{amin2016costbenefit} propose how the installation of a battery storage system along with a PV system in transactive systems might reduce a consumer's electricity bills. In an approach that parallels other transactive PV strategies (e.g., \cite{behboodi2016electric}) they propose to control the charging and discharging cycle of the proposed battery in a transactive system. Using a new cost-benefit analysis method specifically for solar energy systems when combined with batteries, they quantify the economic benefit based on real-time electricity rate and battery cost, to give an exact idea of returns and yearly savings to consumers on their investment. Using real load and generation data they show how an integrated 4kW PV unit reduces the power mismatch between the load demand and solar generation. The result shows that storage-enhanced consumers can maximize their savings considerably on solar investment.

Most recently, Hao et al. \cite{hao2017transactive} expanded on the transactive control approach for commercial building heating, ventilation, and air-conditioning (HVAC) systems for demand response. They describe the system models and  parameters using data collected from a commercial building located on the Pacific Northwest National Laboratory campus. They show how a transactive control market structure works for commercial building HVAC systems, and describe its agent bidding and market clearing strategies. Several case studies are performed in a building controls testbed and calibrated with an EnergyPlus simulation model. They show that the proposed transactive control approach is very effective at peak shaving, load shifting, and strategic energy conservation for commercial building HVAC systems.

\section{Hierarchical Control}
\label{app:hierarchicalcontrol}

This section presents a discussion of how demand response operates in the current hierarchical control of electric power systems and the requirements that transactive control systems must meet to help integrate demand-side resources into system operation.

The electric power industry has undergone a fundamental restructuring over the past 30 years, transforming from a regulated to a market-oriented system. Restructuring has entailed unbundling of vertically integrated organizations into independently managed generation, transmission and distribution systems. As a result, electric power markets have been divided into wholesale and retail systems that interact according to a well-defined, albeit \textit{ad hoc} designs. 

The wholesale power market design proposed by the U.S. Federal Energy Regulatory Commission (FERC) in its April 2003 white paper \cite{ferc2003} encompasses the following core features: (1) central oversight by an independent system operator (ISO) or regional transmission operator (RTO); and (2) a two-settlement system consisting of a day-ahead market supported by a parallel real-time market to ensure continual balancing of supply and demand for power. The objective of an ISO/RTO is to ensure that supply equals demand at every instant, while maintaining system security and reliability and minimizing the total cost of serving the system demand. Optimization is performed on multiple time-scales. The day-ahead settlement system is a pure financial market for generators and load serving entities to create financially binding operating schedules. The real-time energy market allows for the physical exchange of power and addresses deviations between actual real-time conditions and contracted day-ahead agreements. The ISO solves security constrained unit commitment (SCUC) and economic dispatch (SCED) problems in both day-ahead and real-time markets to determine cleared supply and demand, and corresponding locational marginal prices (LMPs), which are reported to market participants. Additionally, to maintain operational balance at any given instant, the ISO runs a balancing reserve market in parallel with the energy markets to calculate the cleared reserve capacities, and the corresponding reservation prices. 

Retail markets have not gone through such a restructuring process. Hence, there is limited participation by distributed assets in wholesale markets through aggregators and no direct participation by smaller assets at all. However, this can be expected to change with accelerated deployment of new ``smart grid'' infrastructure such as digital meters and advanced distribution control systems as were promoted under the Smart Grid Investment Grants starting in 2009. Additionally, FERC Order 755 now requires grid connected short-term storage devices to be treated equitably as conventional generation units when providing regulation services \cite{ferc2011order755}. Similarly, FERC Order 745 requires energy payment of demand response resources at nodal LMPs \cite{ferc2011order745}. As a result a number of wholesale markets now allow distributed assets limited participation in energy markets, usually to meet peak load reduction or emergency services for large-scale demand response programs that serve commercial and industrial users. Feeder level resources still do not participate in wholesale markets, except when provided by demand response aggregators and a limited number of pilot/demo projects, e.g., the Olympic \cite{hammerstrom2007pacific1,hammerstrom2007pacific2}, PowerMatcher \cite{kok2005powermatcher}, Columbus \cite{widergren2014aep}, and Pacific projects \cite{melton2012transactive}. In order to realize the vision of an integration demand response system at the wholesale level, it will be necessary to consider market design changes, the development of more full-functioning system of retail markets, and the integration of the two that provides suitable incentives for wholesale participation by distributed assets. 

There are two key elements to any proposed infrastructure that will facilitate smooth and reliable operations. The first is inter-scale infrastructure that allows devices at various levels to cooperate in determining the efficient allocation of the available resources. The second is the multi-temporal infrastructure that allows devices to shape the allocation they have received within the time horizon in which it is allocated.

The inter-scale infrastructure addresses resource allocation and is used to reconcile supply resource constraints with demand requirement, e.g., feeder constraints versus consumer comfort settings at the retail level. This is accomplished by using real-time prices, such as was demonstrated in the Olympic and Columbus projects. These systems established retail markets that discovered the retail price at which supply equals demand at each feeder in the distribution system, given the current day-ahead prices and prevailing conditions on the feeder and in the homes equipped with price-responsive devices. The Pacific project used a variant of this design for resource allocation that relies on mid-term forecasting usage instead of committing to short-term usage. This system substitutes an index for a price to avoid some of the adverse misconceptions associated with markets in an area that has none at present. This index was developed using a negotiation process that was shown to not converge (see Appendix~\ref{app:pricestability} for one solution to this problem). The Pacific project also differs from Columbus project in the way the formulated signal is presented to the devices.

\begin{figure*}[!t]
	\centerline{
	\xymatrix {
	&	*++[F]{Interconnection}
		\ar@/^/[d]^{f,P}
		\ar@/_0pt/`l[ddl][ddl]^{f}
		\ar@/^0pt/`r[ddr][ddr]^{f}
	\\	 
	&	*++[F]{Area~control}
		\ar@/^/[u]^{\$,Q}
		\ar@/^/[d]^{A,P}
		\ar@/^/[dr]^{A}
	\\	*++[F]{\quad Loads \quad}
		\ar@/^15pt/[uur]^{Q}
		\ar@/_/[r]_{B,\$}
	&	*++[F]{Utility}
		\ar@/^/[u]^{\$}
		\ar@/_/[l]_{P,R}
		\ar@/^/[r]^{P,R,\$}
		\ar@/^/[d]^{A,P,R}
	&	*++[F]{Generators}
		\ar@/_15pt/[uul]_{Q}
		\ar@/^/[l]^{B}
	\\	
	&	*++[F]{Aggregators}
		\ar@{<->}@/^/[u]^{\$}
		\ar@{<->}`r[ur]_>{\$}[ur]
		\ar@{<->}`l[ul]^>{\$}[ul]
	}}
	\caption{Hierarchical Transactive Control System Diagram.}
	\label{fig:transactive_control_system_diagram}
\end{figure*}
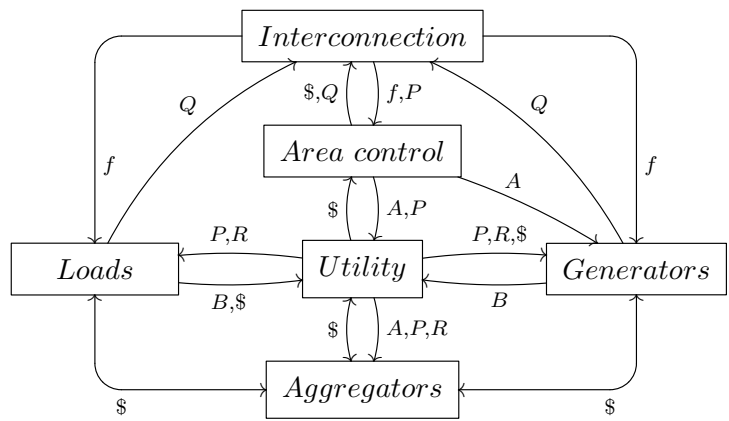
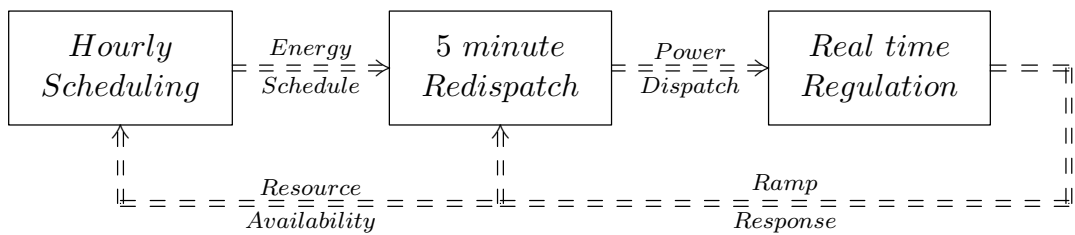
\begin{figure*}[!t]
	\centerline{ \scalebox{1.1}{ \xymatrix {
			*++[F]{\begin{array}{cc}Hourly\\Scheduling\end{array}} 
			\ar@{==>}[rr]^{Energy}_{Schedule}
		&
		&	*++[F]{\begin{array}{cc}5~minute\\Redispatch\end{array}} 
			\ar@{==>}[rr]^{Power}_{Dispatch}
		&
		&	*++[F]{\begin{array}{cc}Real~time\\Regulation\end{array}} 
			\ar@{==>}`r/0pt[rd]`[d]`[ll]_{Ramp}^{Response}[ll]
		&
	\\
		&
		&	\ar@{==>}`l/0pt[llu]_{Resource}^{Availability}[llu]
		&
		&	
		&
	}}}
	\caption{Inter-temporal data flow diagram.}
	\label{fig:temporal_flow}
\end{figure*}

The inter-scale structure is shown in the vertical dimension of \reffig{transactive_control_system_diagram}, while the inter-temporal structure is shown in \reffig{temporal_flow}. At the interconnection level we find the wholesale markets and system operators who set hourly prices and inter-area flow schedules by setting control area import/export schedules that maximize each area's ability to maintain system balance without exceeding interarea flow limits. Control areas establish local prices and respond to deviations in schedules and frequency by sending control signals ($A$) to generators and utilities.  These interact with loads and load aggregators using the local closed-loop bid/response pricing mechanisms implemented at the distribution feeder level.

The real-time and day-ahead market-based feeder and area management systems aim to maximize asset participation at every level in the real-time and day-ahead markets by incorporating the smart grid resources into standard ISO/RTO market structures. This is achieved by solving an optimization problem subject to feeder and area level operational constraints and uncertainties of intermittent renewables and distributed smart grid assets. At the wholesale system level, the ISOs and BAs receive aggregated net load demand, supply bids for distributed assets from the area management systems and generator power supply offers from generators and aggregators. The ISO also runs balancing reserve markets in parallel with the energy and capacity markets to procure reserve energy and capacity to maintain system stability. The cleared or scheduled power setpoints and reserve capacity requirements are then dispatched to the area controllers, which in turn dispatch requirements or price signals to the feeder controllers and distributed assets in their respective retail markets. 

Maintaining operational balance at every instant requires balancing areas to solve an area-wide optimization problem to allocate a portion of their reserve capacity requirements to each feeder based on the resources available. This problem is theoretically amenable to treatment by a market-based process because of the primal-dual nature of the optimization in question.  This process must also maintain adequate area-wide support of frequency and tie-line flows by committing resources to autonomously mitigate deviations in the area control error. Feeders bid resources into the area market and use the cleared prices to dispatch setpoints in real-time to the distributed assets and primary devices needed to meet the performance requirements of the area and feeder control systems. 

At the device level, decentralized control schemes for the distributed assets provide both economic and reliability responses based on self-sensing of frequency, voltage, broadcasts of imbalance signals, current and future prices, and local device conditions. The distributed assets must be dynamically influenced via dispatched control set-points that area reset periodically based on price signals from the retail markets, while responding autonomously and instantaneously to events during contingency operations.

The current standard practice for both direct and indirect control of thermostatic load relies primarily on so-called ``one-shot'' or direct load shedding strategies for emergency peak load relief only. This approach uses a controllable subset of all loads in a particular class, e.g., water heaters or air conditioners, which are transitioned to a curtailed regime that reduces the population average power demand of the end-use category.  After a time, these responsive loads are released and return to their normal operating regimes.  This strategy is known to exhibit fluctuations in aggregate load during the initial response as well as demand recovery rebounds after the loads are released, particularly when thermostatic loads are engaged. To mitigate this behavior, direct load control strategies are sometimes enhanced by either centralized diversification mechanisms, such as using multiple subgroups of the responsive loads dispatched in a sequence that smooths the overall response of the load control system \cite{ruiz2009}, or distributed mechanisms, such as using stochastic control strategies \cite{vrettos2016fast}. Many of these mechanisms require some knowledge of the aggregate thermal response of the buildings in which the loads are operating \cite{perfumo2012}.  To solve the more general tracking problem where load ``follows'' intermittent generation \cite{chu1993}, these mechanisms must address response saturation and loss of diversity \cite{chassin2011}, high sensitivity to modeling errors and noise \cite{elferik2006}, and stability considerations due to feedback delays \cite{bashash2011,roozbehani2010stability}.

Aggregating building thermal load is known to provide a potentially significant resource for balancing purposes \cite{schweppe1980}.  But the hysteresis of standard thermostats requires a switched-mode representation of the individual building thermal response, which in turn gives rise to more complex aggregate load models. The hysteresis also requires that models include so-called ``refractory states''; i.e, states are locked in or out for a certain time after a state transition \cite{chang2013modeling}. These time-locked states are associated with transition delays rather than thermal parameters.  Tractable state space models of aggregate loads can be obtained using model-order reduction strategies that linearize the system model and limit the number of state variables required to represent responsive loads \cite{lu2004,zhang2013}. These models depends on knowledge of the rate at which devices turn on and off and cross the hysteresis band limits, as well as the rate at which the control lockout times expire. Such a state space model minimally represents any thermostat with non-zero deadband. More recent models include the ``battery model'' \cite{hao2013generalized}, which presents aggregate thermostatic loads as simple first-order systems. But these models are intended for small signal response such as frequency regulation, and may not be as useful for sustained peak load curtailment or emergency load shedding.

An alternative thermostatic controller design strategy was proposed to overcome some of the modeling issues arising from the lockout time delays, while not compromising the role of control hysteresis in limiting fast-cycling behavior of thermal loads \cite{chassin2015a}.  This thermostat design uses a discrete-time zero-deadband ($T\Delta_0$) concept that has no refractory states and synchronizes the state transition times with an external signal such as a price coming from a real-time retail double-auction. By using suitably selected sampling rates to limit fast-cycling of equipment $T\Delta_0$ thermostats were hypothesized to give rise to a simple aggregate load model and were shown to have the same comfort and cost savings as conventional thermostats when operated under real-time price tariffs. 

Demand response is becoming a more accepted\footnote{For example, wholesale markets now support demand response in several ISOs and jurisdictions, including PJM, California, and New York.} and important option for utilities to mitigate the intermittency of renewable generation resources \cite{behboodi2016renewable}.   Transactive control is a multi-scale, multi-temporal paradigm that can integrate  wholesale energy, capacity, and regulation markets at the bulk system level with distribution operations \cite{bejestani2014hierarchical}. Under the transactive control paradigm, retail markets for energy, capacity, and regulation services are deployed to provide a parallel realization of wholesale markets at the distribution level. In spite of the conceptual similarity, the behavior of retail markets differs significantly from that of wholesale markets and remains an active area of research \cite{schneider2011analysis}. In particular, load behavior usually dominates the behavior of retail systems, which contrasts with wholesale systems where generation is dominant.  In addition, there are a number of important processes in bulk power interconnection operations that have yet to be integrated fully into the transactive paradigm.  Two very important such processes are system frequency regulation and control area import/export schedule tracking.

Frequency regulation and schedule tracking are jointly regulated using a feedback signal called ``area control error'' or ACE.  The standard mechanism is based on a computation performed in time-domain independently in each control area by evaluating
\begin{equation}
	[ Q(t) - Q_s ] + B [ f(t) - f_s ],
\end{equation}
where $Q(t)$ is the actual net exports from the control area at the time $t$, $Q_s$ is the scheduled net exports, $B$ is the frequency bias, $f(t)$ is the interconnection frequency at the time $t$, and $f_s$ is the nominal or scheduled frequency.  In most realizations the ACE signal is updated by the SCADA system about every 4 seconds and further passed through a smoothing filter so that it changes with a time-constant well in excess of the generating units' fastest response; e.g., 30-90 seconds, with the purpose of reducing wear and tear on generating unit governor motors and turbine valves \cite{kundur1994power}.

Numerous studies examining frequency regulation resource performance using diverse loads have been conducted in recent years. Lakshmanan et al. \cite{lakshmanan2016provision} studied the provision of secondary frequency control in electric power systems based on demand response activation on thermostatically controlled loads in domestic refrigerators in an islanded power system. Observations of household refrigerator response time, ramp rate, and consumer impact show that they provide sufficient fast responsive loads for DR activation, with a typical response time of 24 seconds and a p.u. ramp down rate of 63\% per minute, which can satisfy the requirements for primary frequency control.

Vrettos et al.~\cite{vrettos2016fast} proposed a frequency control scheme designed to augment generation response with load responses, and separate fast-acting load resources from slower responding ones to provide efficient frequency and inter-area power flow using loads.  
\begin{figure*}[!t]
	\centerline{\scalebox{0.95}{\xymatrix{
	&
	&
	&	*++[F]{\alpha} \ar[r]
	&	*++[F]{Loads} \ar`r/0pt[rd][rd]
	&
	\\
	&	*++[F]{Droop} \ar`r/0pt[ru]`[rur][rur] \ar[rr] 
	&
	&	*++[F]{1-\alpha} \ar[r]
	&	*++[F]{Generators} \ar[r]
	&	*+o[F]{+} \ar[dd]
	&
	\\
	&	{\begin{array}{cc}scheduled\\exports\end{array}} \ar[r]^>>>{-}
	&	*+o[F]{+} \ar[d]
	&	{\begin{array}{cc}actual\\exports\end{array}} \ar[l]
	&
	&
	&
	\\
	&	*++[F]{Bias} \ar[r]
	&	*+o[F]{+} \ar[r] \ar`d[rd][rd]
	&	*++[F]{\beta} \ar[r]
	&	*++[F]{Aggregators} \ar[r]
	&	*+o[F]{+} \ar[d]
	\\
	&
	&
	&	*++[F]{1-\beta} \ar[r]
	&	*++[F]{Generators} \ar[r]
	&	*+o[F]{+} \ar[r]^<<<<<{power}
	&	*++[F]{System} \ar`r[rd]`[dllllll]`[llllll]_{frequency}`[ulllll][ulllll] \ar`r[rd]`[dllllll]`[llllll]`[uuulllll][uuulllll]
	&
	\\
	&
	&
	&
	&
	&
	&
	&
	}}}
	\caption{Area control scheme using loads proposed by Vrettos et al. \cite{vrettos2016fast}}
	\label{fig:vrettos2016fast}
\end{figure*}
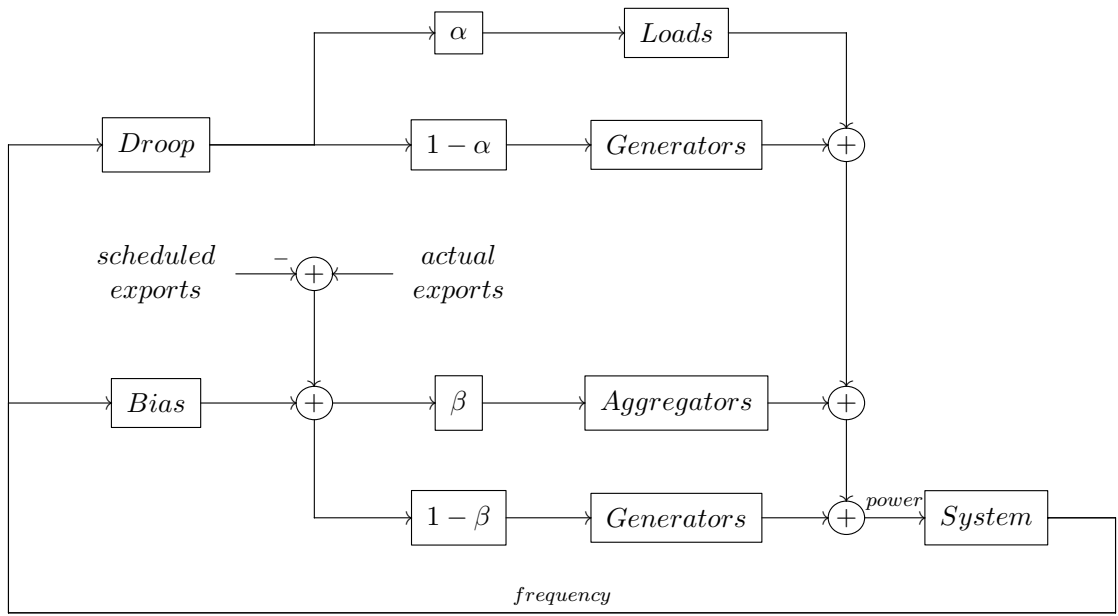
By separating autonomous residential refrigeration loads from dispatchable waterheaters and HVAC systems they achieved the desired response without compromising comfort.  The separation of the loads was based on device characteristics, such as thermal inertia and significant time-dependent power variation, if any. Secondary frequency control is provided by coordinating the runtime of these heterogeneous resources. The solution shown in \reffig{vrettos2016fast} was demonstrated using an optimizer that satisfies the worst-case reserve requirements without violating end-user comfort constraints.  This result provides a useful existence proof, which is significant for any transactive approach that relies on similar mixes of resources but achieves the solution using a distributed system.

Zhong et al. \cite{zhong2014coordinated} developed a large-scale coordination strategy for electric vehicles, battery storage and traditional frequency regulation resources for automatic generation control. Recognizing that response priorities and control strategies for these resources vary with different operating states, they showed that a coordinated control approach not only fully utilizes each resource's advantages but also improves the frequency stability and facilitates the integration of renewable energy.

Falahati et al. \cite{falahati2016grid} examined a model of storage using electric vehicles as moving batteries in deregulated power systems as one way to deal with the frequency regulation problem in a deregulated system with a growing share of intermittent generation resources. They enabled a vehicle-to-grid option for the control of the frequency using an optimized fuzzy controller to manage electric vehicle charging and discharging based on system frequency. The results illustrated satisfactory performance for frequency control of the grid system and verified the effectiveness of the approach for reducing the need for under-frequency load shedding to protect the system against large frequency excursions.

Teng et al. \cite{teng2016challenges} proposed a framework to quantify and evaluate the impact of electric vehicles on island systems like Great Britain. This framework used a simplified power system model to analyze the effect of declining system inertia on primary frequency control and the ability of electric vehicle chargers and batteries to provide resources to mitigate that impact.  Using this model they proposed an advanced stochastic scheduling tool that explicitly models the loss of inertia and assesses the costs and emissions arising from primary frequency control as well as the benefits of having electric vehicles provide primary frequency response. The results of an analysis for Great Britain show that integrating electric vehicles in the primary frequency control system can significantly reduce anticipated cost and emissions growth.

But the results from additional control are not always ideal.  Biel et al. \cite{beil2016frequency} examined the frequency response of commercial HVAC systems by comparing different control strategies for providing frequency regulation demand response.  Aside from noteworthy performance impacts from intra-facility communications delay and control latencies, the authors also observe reductions in energy efficiency when the frequency regulation controls are more active, pointing to the necessity for combined long-term energy cost and short-term regulation response revenue to be considered jointly. In a concurrent study, Khan et al. \cite{khan2016stochastic} followed up on studies by Hao \cite{hao2013generalized} and Sanandaji \cite{sanandaji2016ramping} of the storage-like behavior of thermostatic loads by proposing a stochastic battery model.  This model provides parameters of the battery model and considers changes to the hysteretic thermostatic control in response to frequency. This provides a relatively simple solution to modeling aggregate load control.  However the approach has not been integrated with transactive approaches, and may not account fully for the longer-term endogenous energy conserving integral error feedback that is intrinsic to thermostatic control in general. 

A number of studies of optimal generation control design have been previously reported. Bevrani and Bevrani \cite{bevrani2011pid} studied the general frequency control tuning problem for multiple objectives and proposed three methods for tuning PID controllers to improve the performance of closed-loop systems, including a mixed $\H2/\H\infty$ optimal design method. This approach is easily transferable to a static output feedback control implementation, as is the case with power system area control and any generalized extension where export schedule tracking is desired.  The $\H2$-optimal design method is particularly interesting when there are significant robustness issues to consider, although the authors did not present a solution to the synthesis of an optimal area controller. Jay and Swarmy \cite{jay2016demand} also recently proposed a reinforcement learning-based approach to automatic generation control that was found to minimize frequency deviations by incorporating thermostatic demand response control strategies.

\section{Demand Response in Organized Markets}

FERC issued Order 745 in 2011 to amend its regulations so demand response resources with the capability to balance supply and demand can participate in organized wholesale energy markets as an alternative to generation resources. The order introduced requirements that (1) dispatched demand response resources satisfy a net-benefit test, and (2) demand response resources are compensated for the services they provide to the energy market at the locational marginal price (LMP). This approach for compensating demand response resources was intended to ``ensure the competitiveness of organized wholesale energy markets and remove barriers to the participation of demand response resources, thus ensuring just and reasonable wholesale rates'' \cite{ferc2011order745}. 

Critics of Order 745 have pointed out that demand response is essentially unlike generation because it is exercised as a call option on the spot energy market, the value of which is the LMP minus the strike price, which in the case of retail consumers is the tariff rate \cite{hogan2009}.  Others contend that the value of demand response is the marginal forgone retail rate \cite{borlick2010}. However it is valued, the question remains whether FERC Order 745 effectively guarantees double compensation for demand response by providing responsive load both the cost savings from energy not provided by the retailer and an LMP payment for not using same increment of energy. Such a signal might lead a firm to halt operations even though the marginal benefit of consuming electricity exceeds the marginal cost at LMP. In his comments to the Commission during prosecution of the order Hogan argues that ``the ideal and economically efficient solution regarding demand response compensation is to implement retail real-time pricing at the LMP, thereby eliminating the need for [wholesale] demand response [compensation]'' \cite{hogan2009}. 

These arguments are largely academic if demand response cannot be deployed broadly for technical reasons.  To resolve the technical questions regarding the large-scale feasibility of demand response in the context of smart grid the Olympic and Columbus demonstration studies were conducted in 2006--2007 and 2013, respectively. The objective of both studies was to address the technical questions regarding the so-called ``price-to-devices'' challenge \cite{smith2005} by demonstrating the transactive control approach to integrating small-scale electric equipment with utility electric power distribution system operations as a first step toward integrating distributed generation and demand response into wholesale operations. Transactive control thus refers to a distributed resource allocation strategy that engages both electricity suppliers and consumers using market-based mechanisms at the retail level for the purpose of enabling demand response by the utilities at the wholesale level \cite{hausman2004}.

A number of problems were identified following the Olympic and Columbus demonstrations. The most significant among these were the following.

\begin{description}

\item[Short-term Price Stability] When demand response resources become scarce, price volatility increased. This creates an undamped feedback that resulted in rail-to-rail actuation of the demand response resources. The oscillation develop because resource state diversity is defeated by the larger price signals, which in turn drives large price volatility. The individual resources themselves are not adversely affected by these undamped oscillations. But from the system perspective this is regarded as an undesirable regime.  The oscillations continue until either sufficient demand resources are returned online to restore state diversity that overcomes the oscillations, or the resource constraint that drove the price surge goes away.

\item[Lack of Forward Prices] The expectation price and price volatility should be obtained from forward markets. In the demonstrations they were developed from historical responses of the system.  This created a form of delayed feedback that could give rise to long term price instability.

\item[Lack of Multi-level Market Integration] The retail markets should not be operated autonomously or in open loop. Although the wholesale prices were used in formulating the feeder price bids, at no point was the retail market used to formulate bid or forecasts for the wholesale market.

\item[Lack of Integration With Regulation System] The retail market-controlled resources should be used to arm and reset regulation services, such as fast-acting demand response with autonomous controls.  These resources should be controlled by adjusting setpoints based on resource needed at the system level.  For example, grid-friendly appliance controllers were set to shed load at 59.95 Hz with a probability of 1.0. This resulted in a fixed aggregate control gain that could have been adjusted based on system need during each dispatch period.

\end{description}

Despite of these problems, the demonstration projects were largely considered technical successes. In fact, it can be argued that our understanding of these issues is in large part due to the level of complexity attempted by the projects.

Without aggregate load control, introducing price-responsive demand requires engaging millions of very small participants in the unit-commitment and economic dispatch process. This may be impractical due to the computational complexity of the process just for the thousands of larger suppliers already involved. Strategies for addressing this challenge generally involve aggregation at the retail level that enables the integration of demand units by proxy of a reduced number of larger representative units \cite{vrettos2016fast}.

Using markets to solve electricity resource allocation problems at the wholesale bulk system level is well-understood \cite{stoft2002} . But transactive control takes the idea to the retail level by enabling the resource allocation problem at the distribution level first before integrating it at the wholesale level. These markets are designed to find a Pareto-optimal allocation of distribution capacity, distributed resource and demand response to resolve how much wholesale resource is required, and determine how much distributed generators produce, and customers consume in the coming time interval.  The transactive control systems demonstrated use distribution capacity markets to determine the energy price that minimizes the imbalance between supply and demand for electricity by participating equipment during the next operating interval \cite{chassin2014electric}.  The system computes a 5-minute retail real-time price (RTP) that reflects the underlying wholesale locational marginal price (LMP) plus all the other distribution costs and scarcity rents arising from distribution capacity constraints. The real-time price comes under a new tariff designed to be revenue neutral \textit{in the absence} of demand response. 

Distributed generation, load shifting, curtailment and recovery are all induced by variations in real-time prices. In doing so, the transactive control system can reduce the exposure of the consumers and the utility to price volatility in the wholesale market and the costs of congestion on the distribution system \cite{chassin2015optimization}. Retail prices are discovered using a feeder capacity double auction that can be used to directly manage distribution, transmission or bulk generation level constraints, if any. Distributed generation is dispatched based on consumers' preferences, which they enter into an advanced thermostat that acts as an automated agent bidding for electricity on their behalf.  Thermostats both bid for the electricity and modulate consumption in response to the market clearing price. By integrating this response with a price signal that reflects anticipated scarcity the system closes the loop on energy consumption and can improve resource allocation efficiency by ensuring that consumers who value the power most are served first. At the same time, consumers provide valuable services to the wholesale bulk power system and reduced energy costs at times of day when they express preferences for savings over comfort.

In the California ISO, demand response is acquired using two different products, Proxy Demand Resource (PDR) and Reliability Demand Response Resource (RDRR) \cite{caiso2017demand}.  PDR bids supply in CAISO energy, non-spin and residual commitment markets for economic day-ahead and real-time dispatch.  RDRR bids supply for reliability purposes as energy services only and dispatches economic day-ahead and reliability real-time.

In the eastern interconnection, PJM obtains demand response services from Curtailment Service Providers (CSP). CSPs are entities responsible for demand response activities in PJM wholesale markets \cite{pjm2016demand}.  These can be companies that focus exclusively on customer demand response capabilities, or they can be utilities, energy service providers, or other types of companies that can provide the necessary services to PJM.  PJM uses these services to obtain two kinds of demand response, economic and emergency response. Customers can participate in either or both, depending on circumstances.  

PJM treats emergency demand response as a mandatory commitment which is dispatched similarly to generation when reliability resources are in short supply or under emergency operating conditions.  Four products are generally used: (1) Limited DR for no more than 6 hours a day on up to 10 days during summer months, (2) Extended Summer DR for no more than 10 hours a day for up to 10 days during extended summer months, (3) Annual DR for no more than 15 hours a day for any day of the year, and (4) Base DR for up to 10 hours any day of the year. Pricing is driven by the capacity market as defined under the Reliability Pricing Model (RPM), with payments made monthly for availability during expected emergency conditions.

In contrast, PJM economic demand response is a voluntary commitment to reduce load in the energy market when wholesale prices are higher than the monthly net benefits price at which the benefit of reduction exceeds the cost of paying for the demand response. These resources are used to displace generation and PJM expects these resources to reduce load measurably, and impose penalties for loads that fail to deliver the expected load reductions.  

Economic demand response resources can also provide Ancillary Services to the wholesale market if enabled with the appropriate infrastructure.  These include (1) Synchronized Reserves that act within 10 minutes of dispatch, (2) Day-Ahead Scheduling Reserves that act within 30 minutes of dispatch, and (3) Regulation that follow PJM's regulation and frequency response signal.  Participating in the Ancillary Services market is voluntary, but once cleared in the market, performance is mandatory.

Demand response programs in New York (NYISO) and Texas (ERCOT) have many similar characteristics and products. But the specifics of their designs only further demonstrate how no two wholesale markets use precisely the same rules and definitions. This only exacerbates the existing Balkanization of the North American energy market designs and presents yet another barrier to the widespread adoption of demand response as a technology to mitigate the reliability impacts of renewable intermittency.

\subsection{Retail Competition}

Retail competition was introduced in Europe, Australia, New Zealand and some US states over the last two decades.  The theoretical rationale for retail competition goes against conventional wisdom in retailing because electricity is not a storable good and the infrastructure needed for retail delivery is largely a natural monopoly. Furthermore, the homogeneity of electricity limits opportunities for potentially profitable retailing activities such as presentation, packaging, bundling, and co-branding. These combine to limit the benefit of introducing retail competition in three signicant ways \cite{defeuilley2009retail}.

\begin{enumerate}

\item The potential demand for a supplier to meet is limited by the low revenues that retailing activities can generate.

\item Retail markets must be created where none exist today, which requires consumers to engage in searching, learning, and transaction activities that are costly barriers to switching from the incumbent utility.

\item The homogeneity of the products limits the differentiation and discrimination necessary to create value-added services for the consumer.

\end{enumerate}

The expectation was that entrepreneurial innovation would provide social benefits and would overcome these barriers. Retail competition was expected to reduce market imperfections \cite{waddams2004spoilt}, motivate the discovery of new products and services \cite{littlechild2002competition}, stimulate customer awareness, and drive competition in generation \cite{littlechild2000we}. But the results seem not to be supported by the data observed initially. One key indicator is that the percentage of customers active in the market remains quite low: no retail market reports more than 50\% participation and a majority reported less than 10\% by 2009 \cite{defeuilley2009retail}. Moreover, switching costs have not decreased over time, suggesting that learning effects are not intensifying and customers are not reducing the risks and uncertainties in their decisions. In Great Britain and Norway, for example, the costs of switching did not fall below 10\% by 2009 \cite{von2009electricity}, while the rate of customer switching remains high \cite{pollitt2014dismantling} compared to other markets.

In California another form of retailing has emerged called ``Community Choice Aggregation'' or CCA \cite{faulkner2010community}. CCAs allow cities and counties to combine the demand of electricity customers in their jurisdictions to procure electricity through bilateral contracts or markets.  The advantage a CCA has relative to a regulated investor-owned utility (IOU) is threefold. First this gives the community served more control over the energy sources used and allows the CCA to market a differentiated product (e.g., renewable energy) that the incumbent IOU cannot provide due to regulatory constraints. Second the CCA may not be required to purchase the same level of ``firm'' resources, which may be a significant cost-consideration for renewable energy sources.  Finally, a CCA presents consumers with some choice in purchasing electricity under a cooperative structure, in spite of the fact that it is not individual choice in the manner envisioned by retail competition advocates. 

The result of retail competition and community aggregation overall is mixed. There is an ongoing segmentation of the retail market into participating and non-participating consumers, which is undermining any persistent generalization of retail competition and community choice in electricity delivery. Regulators are justifiably concerned that underserved and low-income communities are not well-served by retail competition and community choice, even when they choose not to adopt it for themselves, partly because as those who can afford to flee the regulated utility environment do so, those who remain behind are left with most of the costs that are not recovered by these new solutions. Nonetheless, the incumbent utilities still retain significant ``brand loyalty'' where retail competition has been adopted, and for those consumers who are unable or unwilling to participate it is not at all obvious that they have received much benefit from the transformation of the retail marketplace.

\section{Scheduling, Dispatch and Regulation}

Responsibility for the reliability of electricity interconnections is shared by all the operating entities within each interconnection. In a traditional power system, these entities are vertically integrated. A committee process involving all the entities within each power pool establishes the reliability criteria utilities used for planning and operations.  Typically, the operating entities belong to larger regional coordinating councils so that they can coordinate their criteria with neighboring power pools.  Since 1965 these regional councils have been organized under what is now called the North America Electric Reliability Corporation (NERC), which establishes the recommended standards for system reliability \cite{nerc2017standards}.

With the evolution toward market-based operations in recent decades, vertically integrated operating entities have been broken up into generation companies (GENCOs), transmission owners (TRANSCOs), load serving entities (LSEs), and energy traders that do not own assets, all of whom are collectively the market participants \cite{bohn1988spot}. The responsibility for ensuring the reliability of a control area is delegated to independent system operators (ISO) or regional transmission operators (RTO).  In general market participants have a duty to provide accurate data about their assets and costs as well as follow the dispatch orders of the ISO/RTO.  The ISO/RTO has the duty to ensure that each market participant meets the reliability rules and determines the dispatch orders necessary for the electricity supply and demand to match according to NERC's reliability standards. This system is predicated on a successful competitive market in which private decentralized trading and investment design work to allow substantial commercial freedom for buyers, sellers and various other types of traders \cite{hogan1998primer}.

The method used to implement such a system planning and operating model employs a two-stage process referred to as the unbundled or two-settlement approach:

\begin{enumerate}

\item Unit-commitment (UC) is a days-ahead process that determines the hourly operating set points of the generation assets based on their bid energy prices and the forecast system load.

\item Economic-dispatch (ED) is an hours-ahead process that determines the real-time generation schedules and procures additional supply to ensure system reliability.

\end{enumerate}
This approach can be used for both regulated and unregulated markets and the analysis method is similar for both short-term operations and long-term planning with only the caveat being that ISOs must perform the system studies for deregulated markets to determine whether additional generation and transmission may be required. 

Overall the timeframes for planning and operations can be separated into the following security functions \cite{chow2005electricity}:

\begin{enumerate}

\item	Long term planning (2-5 years) determines needed investments in generation and transmission.

\item	Resource adequacy (3-6 months) secures generation to serve expected load and sets long-term maintenance schedules.

\item	Operation planning (1-2 weeks) coordinates short-term maintenance schedules and long-lead generation.

\item	Day-ahead scheduling (12-24 hours) performs a security-constrained UC using energy bids.

\item	Real-time commitment and dispatch (5-180 minutes) performs real-time security-based economic balancing of generation and load.

\end{enumerate}

For time-intervals shorter than about 5 minutes, the reliability of the system is delegated entirely to the generation and loads according to reliability standards promulgated by NERC and coordinated separately by each interconnection.

Modern bulk electric interconnections are constrained by the physical requirement that electric energy is not stored in any substantial way during system operations.  Historically, utility operations have focused on controlling generation to ``follow'' load and ensure that at every moment supply exactly matches demand and losses. To make electric utility planning and operation economical and manageable, the industry divides generation resources into three principal categories: base load, intermediate load, and peak load \cite{kundur1994power}.  

Base load generation is the bottom portion of the supply stack that essentially runs uninterrupted throughout the year (except during maintenance or unplanned outages). Intermediate generation runs continuously but only seasonally as the diurnal load nadir rises and falls.  Peak generation is the supply that must be started and stopped daily to follow the diurnal load variations.  Each of these types of generation also provides regulation and reliability resources to help control frequency and respond to contingencies and emergencies in generation and transmission operations.  

For decades load had not generally been considered part of the overall planning and operations model of electric interconnections except to the extent that its growth set the conditions for capacity planning.  But in recent years increasing thought has been given to the role that load can play as a demand resource that (1) reduces the need for new conventional generation resources, (2) avoids using generation resources in inefficient ways, and (3) enables the addition of generation resources that exhibit substandard performance characteristics when operating under the conventional load-following paradigm \cite{kundur1994power}.

Today the term ``demand resource'' encompasses a wide range of products, services and capabilities related to the control and management of load in electric systems.  Prior to the advent of ``smart grid'' technology, demand resources were primarily considered for planning purposes, such as demand-side management (DSM) programs, and very limited operational purposes such as \textit{in extremis} under-frequency or under-voltage load shedding programs (UFLS/UVLS) \cite{glover2012power}.  DSM programs are planning programs that focus on energy efficiency and other long-term demand management strategies to reduce load growth so that the need for significant new generation capacity investments can be deferred or eliminated.  Generally these programs pay for themselves by reducing capital costs for a number of years, possibly indefinitely.  DSM programs helped the industry transition from its pre-1970s 7\% annual capacity growth to the sub-3\% growth prevalent today in modern electricity interconnections. 

But DSM programs have a number of long-term limitations that prevent their application to other system planning or operations objectives.  First, energy efficiency is generally a diminishing return because every additional dollar invested replaces less inefficient load.  In addition, DSM programs can give utilities an incentive to substitute investments in a few larger (presumably more efficient) base load units with numerous smaller (generally less efficient) intermediate units or even (typically very inefficient) peaker units.  Finally, DSM programs generally do not provide the capabilities and controllability needed to address some recent new planning and operations challenges such as generation intermittency, the lack of transmission capacity investments, evolving load characteristics, new ancillary service market designs and short-term/real-time energy price volatility \cite{eto2002innovative}.  

In contrast, UFLS and UVLS are strictly operations programs that focus on very short-term load curtailments under severe contingencies. They are used when all or part of the electric interconnection is threatened by a large unexpected loss of generation or system separation that creates a power imbalance which can only be remedied by drastic and immediate reductions in load.

Load shedding programs also have important limitations because they are pre-programmed actions armed to respond to specific circumstances identified during planning studies. They are not the flexible and graduated responses needed for more general regulation and balancing operations.  Load shedding programs also tend to indiscriminately disconnect loads and do not have the ability to affect only less critical end-uses such as air-conditioners and water-heaters.  

As intermittent generation becomes a standard element of the generation fleet, the interest in using demand response as a substitute for new controllable generation can be expected to grow.  In addition, demand response has long been regarded as necessary because it reveals the elasticity of demand in ways that mitigate supply-side market power.  

\begin{figure*}[!t]
	\includegraphics[width=\textwidth]{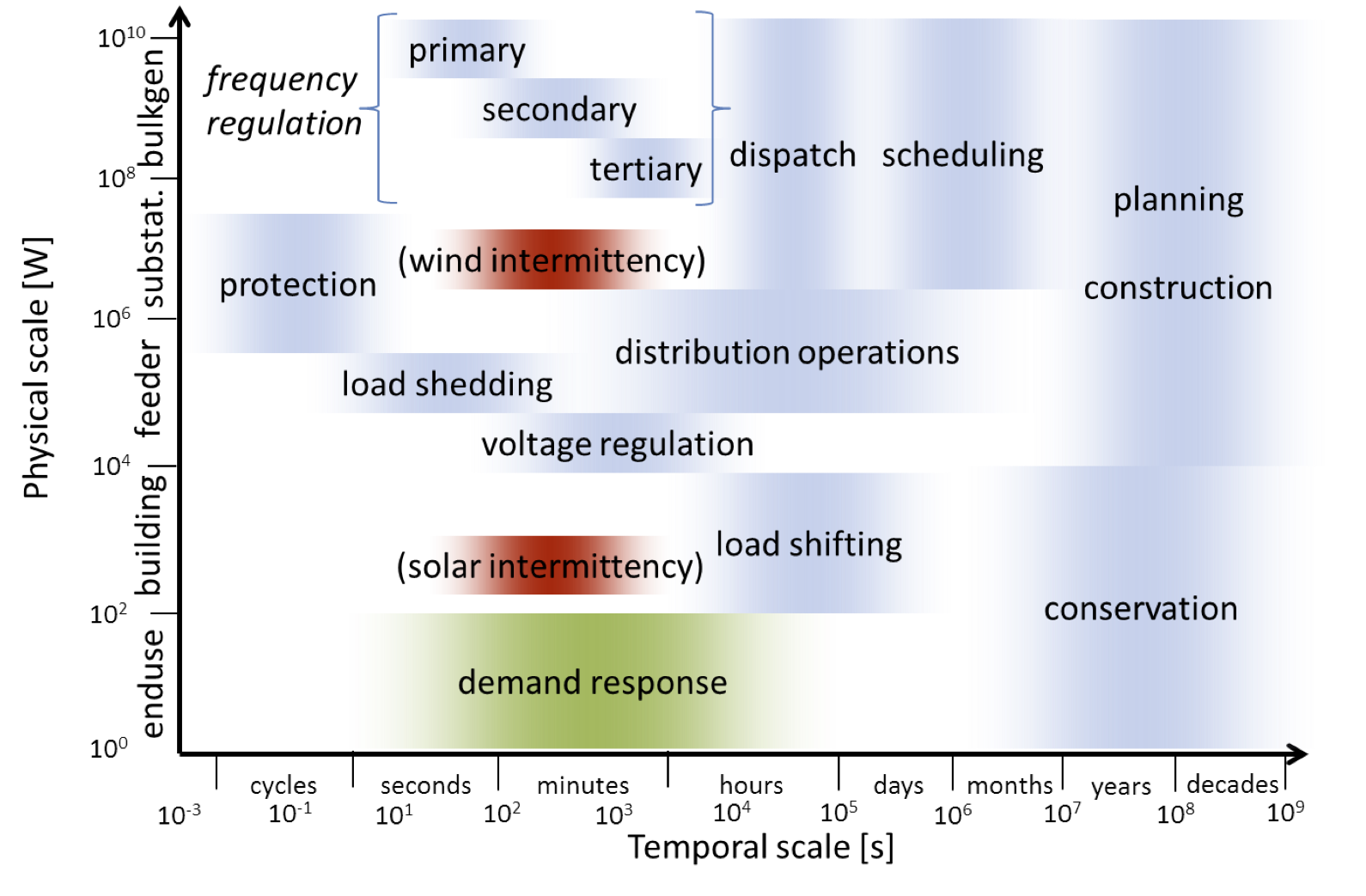}
	\caption{Control and resource variability physical and temporal scales}
	\label{fig:intermittency_scales}
\end{figure*}

But the mismatch in the characteristic size of loads, their timing, and their uncertainty relative to conventional generation is a significant obstacle to using demand response to simultaneously displace supply-centric reliability services and mitigating generator market power: there are relatively few easily observed generators and their characteristic response times are relatively slow compared to overall system dynamics. Loads in contrast are far smaller, far more numerous, and for more difficult to observe. But they are potentially far faster acting than the overall system dynamics \cite{kundur1994power}. The physical and temporal scales of resource variations are shown in \reffig{intermittency_scales}, and we can see where demand response and renewable generation intermittency time scales match well, while the physical scale does not.

Bulk power system planning, operation, and control have generally been designed to consider the characteristics of generators and treated loads as a ``noisy'' boundary condition.  Thus load control remains quite difficult to incorporate into bulk system planning and operation. In general, the approach to addressing this fundamental mismatch is to devise demand aggregation strategies that collect numerous small fast acting devices with high individual uncertainty into a few large slower acting aggregations with reduced uncertainty. While not requiring every electric customer to participate in wholesale markets, demand aggregation provides a means of increasing consumer participation in system resource allocation strategies market-based or centrally controlled and thus can mitigate price volatility whether for energy, capacity, or ramping services \cite{siddiqui2003price}. 

From an economic perspective, aggregating electricity customers can be viewed as a means of capturing consumer surplus to increase producer surplus, by segregating consumers into groups with different willingness to pay.  Three general approaches are usually employed to creating consumer aggregation for either operational or economic objectives:

\begin{enumerate}

\item Economic aggregation is achieved using price discrimination methods such as different rates for different customer classes, product differentiation, and product or service bundling strategies.

\item Social aggregation is achieved using various subsidy programs, and other social group identification strategies such as environmental, green or early-adopter programs.

\item Technical aggregation creates technical structures that either directly aggregate consumers, or indirectly enable economic or social aggregation.  Technical aggregation can be accomplished using service aggregators, creating technological lock-in with high barriers to entry or exit, or constructing local retail markets independent of wholesale energy, capacity, and ancillary service markets.

\end{enumerate}

Variable or intermittent generation is a growing fraction of the resource base for bulk power systems.  The variable character of certain renewable resources in particular is thought to undermine the overall reliability of the system insofar as forecasts of wind and solar generation output have greater uncertainty than more conventional fossil, nuclear or hydroelectric generation resources, or even load. As a result, the expectation is that while variable renewable generation resources do displace the energy production capacity of fossil power plants, they may not displace as much of the power or ramping capacity of those plants.  Consequently, the variable nature of renewable resources may indicate that they do not offer as much emissions benefit as expected if one were to assess their impact simply on energy production capacity \cite{heinberg2016renewable}. 

It seems intuitive that demand response should be able to mitigate the capacity and ramping impacts of variable generation by reducing the need to build and commit fossil generation to substitute for reserves or ramp in place of fast-changing renewable generation.  But this substitutability is constrained by (1) the nature of variable generation, the role of forecasting, and the impact of resource variability on the emissions and economics of renewable resources; (2) the nature of load variability and how demand response is related to load variability; and (3) the characteristics of end-use demand and the impact of demand response on energy consumption, peak power and ramping rates over the various time horizons that are relevant to the variable generation question.  

Taken together, these constraints and interactions provide the basis for assessing the economic and environmental impacts of controllable load and demand response resources on various time scales. It is by virtue of the downward substitutability of reserve resources that we can assume that the variability impact of renewable generation is exactly the opposite and always less than the benefit of the same controllability in demand response, and we can assess the value of demand response using this inequality as a guide.

Taken as a whole, this assessment of the current situation and the benefits of demand resources for energy, capacity, and ramping response provide a basis to justify a major investment of load as a resource and the infrastructure needed to integrate it with the system as a whole.  

\subsection{Technical Requirements}

The remainder of this section focuses on the technical requirements to realize a broad vision of what is now called more generally ``Transactive Energy'', which is based on the original transactive control concept but encompasses the entire energy system. The approach is based on a multitude of market-like mechanisms used to discover the prices that most efficiently schedule, dispatch, and regulate system resources. The concept is proposed as the long-term solution to the challenge of transforming the power grid into a 21$\mathrm{^{st}}$ century system that meets 21$\mathrm{^{st}}$ century needs.

What is missed by most of the existing literature is the strong connection between scheduling, dispatch and regulation on the one hand, and the degeneracy of energy, power and ramp control\footnote{Control degeneracy is the recognition that one cannot independently control the energy use, power level and ramp rate because each is the derivative of the previous.}  on the other hand. Having recognized this problem, the designers of transactive energy infrastructures suggest ten key elements required to successfully implement a complete transactive control system. The breadth of these requirements reflects not only the high complexity and difficulty of the overall design problem, but also the broad scope of the concept itself.  These requirements cover the following processes:

\begin{enumerate}

\item Incentive policies and market mechanisms;

\item Market-based services;

\item Device-level controls;

\item Retail-wholesale integration;

\item Inter-temporal coordination;

\item Balancing services;

\item Balancing area control;

\item Feeder control;

\item Integrated autonomous and market operation; and

\item Off-normal operation.

\end{enumerate}

\subsubsection{Incentive Policies and Market Mechanisms}

Control policies and market mechanisms should provide the correct incentives for the full range of distributed assets. These assets must be able to express their ability, willingness, and desire to modify their supply or demand and respond to the system's overall conditions as they change over time. These policies and markets must lead to the discovery of prices and provide incentives for participants to share accurate data that leads to a coordinated device response and precisely meets the needs of the grid, as a function of time and location, from the lowest-cost resources available. More specifically, a range of incentives and resulting bidding strategies must align with operational and capital costs, applicable in both vertically-integrated and restructured market environments, to ensure appropriate levels of customer participation. This must take into account the utility revenues needed to justify the investment in, and operation of the interconnection as a collection of independent financial and operational entities with both cooperative operational goals and competing financial goals. 

\subsubsection{Market-based Services} 

Current market structures do not support a level playing field for distributed assets when compared to conventional generation. The proposed paradigm seeks to create an equitable market mechanism for coordinating and controlling all system assets through a distributed, self-organizing control paradigm that protects customer choice but encourages and coordinates participation. This is the purpose of the so-called ``transactive'' paradigm. Distributed smart grid asset participation in the wholesale market must be coordinated through a hierarchical architecture of nested market mechanisms. This requires the design of retail markets, but leaves the actual functional control at the device level. This also allows load-serving entities (LSEs) to play their natural role as a resource aggregator in the retail markets and paves the way for independent third party aggregators to develop optimal portfolios to sell to the utilities.

This does not necessitate complete structural changes to current ISO/RTO day-ahead and real-time structures (system level). Rather it complements them by providing institutional mechanisms that integrate retail and wholesale markets using continuous feedback controls. At each of the hierarchical levels (feeder and area), available resources – whether demand or supply, whether energy, capacity, or ancillary services – are aggregated from the level below while considering local constraints, such as energy allocations, capacity limits and ramping reserves. Device level bids are aggregated by feeder level management systems while applying local constraints by clearing retail capacity markets such as those demonstrated in the Olympic and Columbus projects. 

The feeder level bids are then cleared by the area level markets, which submit aggregated or residual bids into the ISO/RTO wholesale market. Conversely, the area and feeder markets then receive the cleared price and dispatch quantities from the ISO/RTO, which are eventually passed down to the end-use customers. This forms a feedback mechanism for a closed-loop, multi-level optimization problem that engages distributed assets in the wholesale market. 

The same structural formulation is applied in both ahead and real-time markets. The only difference between ahead and real-time markets is the formulation of agents' optimization problems and the source of the information needed to formulate bids. When using forward markets, each resource bids using residual allocations of quantities from longer-term markets.

\subsubsection{Device-level Controls}

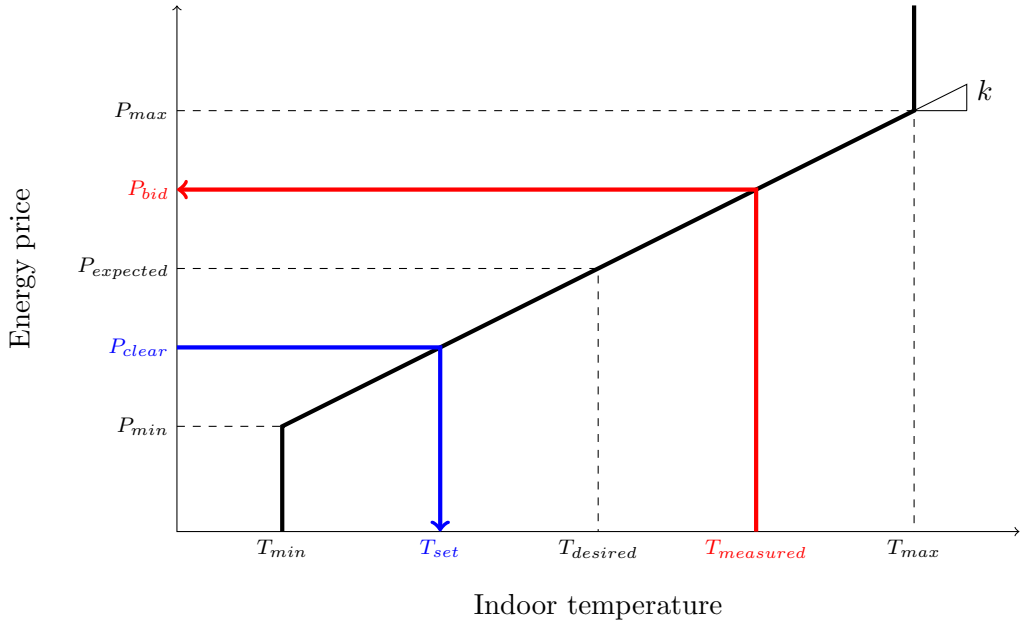
\begin{figure*}[!t]
	\centering
	\begin{tikzpicture}[scale=7]
		\draw [<->] (0,1)--(0,0)--(1.6,0);
		\node [below] at (0.8,-0.1) {Indoor temperature};
		\node [above, rotate=90] at (-0.25,0.5) {Energy price};
		\draw [ultra thick,black] (0.2,0)--(0.2,0.2)--(1.4,0.8)--(1.4,1.0);
		\draw [black] (1.5,0.8)--(1.4,0.8)--(1.5,0.85)--(1.5,0.8);
		\node [above right] at (1.5,0.8) {$k$};
		\scriptsize 
		\draw [dashed,black] (0,0.8)--(1.4,0.8)--(1.4,0);
		\node [left] at (0,0.8) {$P_{max}$};
		\node [below] at (1.4,0) {$T_{max}$};
		\draw [dashed,black] (0,0.2)--(0.2,0.2);
		\node [below] at (0.2,0) {$T_{min}$};
		\node [left] at (0,0.2) {$P_{min}$};
		\draw [dashed,black] (0,0.5)--(0.8,0.5)--(0.8,0);
		\node [left] at (0,0.5) {$P_{expected}$};
		\node [below] at (0.8,0) {$T_{desired}$};
		\draw [->,ultra thick,red] (1.1,0)--(1.1,0.65)--(0,0.65);
		\node [below,red] at (1.1,0) {$T_{measured}$};
		\node [left,red] at (0,0.65) {$P_{bid}$};
		\draw [->,ultra thick,blue] (0,0.35)--(0.5,0.35)--(0.5,0);
		\node [left,blue] at (0,0.35) {$P_{clear}$};
		\node [below,blue] at (0.5,0) {$T_{set}$};
	\end{tikzpicture}
	\caption{Thermostatic device control with price-based bid feedback for a cooling regime using comfort control $k$.}
	\label{fig:transactive_thermostat}
\end{figure*}

At the lowest level devices use price and other information to autonomously determine appropriate actions and apply their own constraints to local control processes. In the Olympic and Columbus demonstrations of this approach HVAC loads responded to changes in normalized prices by adjusting the thermostat set-point utilizing smart thermostat and smart-meter technology, as illustrated in \reffig{transactive_thermostat}. 
These devices bid the price point for their on/off decision as well their power quantities into a retail market.  The price point is a function of the difference between the desired air temperature and the current air temperature and the quantity a function of recent metering. Customers are actively engaged with a simple user interface that allows them to choose how much demand response they provide from a range between ``more comfort'' and ``more savings'' with a simple slider control. This parameter $k$ allows consumers to determine the level of market participation. They can always override the response by either changing the bid response curve or removing the device from the market altogether, provided they are willing to pay potentially higher prices were they to occur. This approach protects customer choice, while continuously rewarding participation.  The Olympic project extended this concept to commercial VAV control units, as well as municipal water pumping facilities, and various types of distributed generating units.

Similar device bid and response mechanisms must be created for other distributed assets, including distributed storage, distributed generation, and smart appliances. The US Department of Energy's Office of Electricity and General Electric Appliances \cite{fuller2013modeling} showed the benefits of multi-objective controls for distributed assets for a wide range of devices. The end result would be an environment and a set of rules for participation where vendors can create additional bidding and control strategies, depending upon the goals of the customer, ranging from relatively simple to highly complex optimization routines or predictive algorithms. Design of device level controls and bidding strategies forms the basis for their participation in retail markets. Equitable treatment of distributed assets in the wholesale markets is accomplished through retail-wholesale integration as described in the following section.

At the device level, distributed assets should provide multiple services at different time scales: (1) respond to market prices both ahead and real-time, (2) respond to imbalance signals, and (3) respond autonomously to reliability needs inferred from ambient frequency and voltage signals.

Autonomous responses are critical for many reliability purposes where there may not be time to communicate needed actions through a wide-area network. Appliance and equipment manufacturers are rapidly moving toward mass production of devices with smart grid capabilities that can be leveraged for this purpose. However, utilities and balancing authorities are hesitant to support such deployments, because the response of fleets of such devices has not been fully integrated with their control schemes for grid stability. Distributed assets must be equipped with autonomous controllers and include settings to arm them according to instructions from feeder, area or system levels. In this way, the autonomous immediate response of devices can be continuously tailored to system needs, such as low system inertia due to high on-line renewable generation and inverter-based loads.

To provide the multiple services, at the device level, distributed assets must be equipped with multi-objective control strategies designed to enable single resources to provide multiple benefits to the system. These control strategies must be accounted for in the coordination problem, for example, by using receding horizon optimization techniques (viz. model predictive control). Device level controls introduced in previous demonstrations must be expanded by including similar easy-to-use (from the customer perspective), economically driven responses for all smart grid assets, including other smart appliances, storage, and distributed generation to provide cost-effective, controllable solutions.

\subsubsection{Retail-Wholesale Integration}

One of the main objectives of this paradigm is to offer a comprehensive framework that fully integrates retail and wholesale power markets. This framework must provide a way for end-users (distributed assets) to contribute indirectly in wholesale markets. 

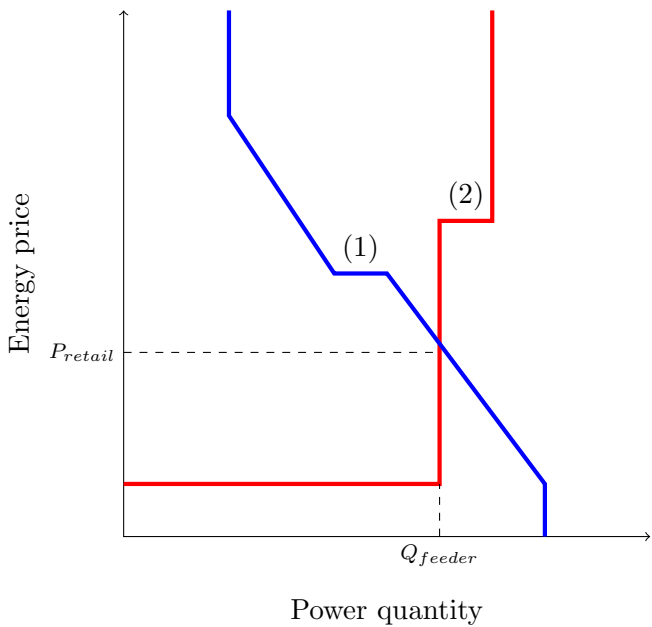
\begin{figure}[!t]
	\centering
	\begin{tikzpicture}[scale=7]
		\draw [<->] (0,1)--(0,0)--(1,0);
		\node [above,rotate=90] at (-0.15,0.5) {Energy price};
		\node [below] at (0.5,-0.1) {Power quantity};
		\draw [ultra thick,red] (0,0.1)--(0.6,0.1)--(0.6,0.6)--(0.7,0.6)--(0.7,1.0);
		\node [above] at (0.45,0.5) {(1)};
		\draw [ultra thick,blue] (0.2,1.0)--(0.2,0.8)--(0.4,0.5)--(0.5,0.5)--(0.8,0.1)--(0.8,0.0);
		\node [above] at (0.65,0.6) {(2)};
		\draw [dashed,black] (0,0.35)--(0.6,0.35);
		\scriptsize
		\node [left] at (0,0.35) {$P_{retail}$};
		\draw [dashed,black] (0.6,0)--(0.6,0.1);
		\node [below] at (0.6,0) {$Q_{feeder}$};
	\end{tikzpicture}
	\caption{Storage resource allocation using retail markets. The battery charge bid (1) is accepted while its discharge bid (2) is not, even though the feeder appears congested.}
	\label{fig:retail_market}
\end{figure}

Retail market designs must not only facilitate interactions between end-users (distributed assets) and the feeder level management system. The feeder level management system must coordinate the behaviors of the distributed assets within their respective retail markets, as well as consolidate the net offering for area and wholesale markets. This provides an avenue to inject local constraints, which are often overlooked when solving system-wide problems using distributed resources. Feeder level optimization and control for a real-time retail market was demonstrated during the Olympic and Columbus projects. In both of these projects, system-wide constraints (in the form of wholesale market prices or LMPs) are coupled with local constraints (local feeder capacity) to clear retail markets and provide both local and system-wide benefits using demand response resources, as illustrated in \reffig{retail_market}. 
Effectively, the system enables customers to reduce their energy consumption during high price events to reduce energy costs, while coordinating devices responses during localized congestion events to decrease demand and deploy local resources, providing a system for equitably rewarding customers for participation. Distributed generation and storage similarly bid into the retail market, subject to run time constraints (e.g., a maximum number of allowable run hours). While successfully showing that distributed assets can participate in retail level markets, the distributed assets in these projects did not affect the wholesale price---they only reacted to wholesale prices and local constraints. 

The feedback loop is closed by integrating retail markets through to wholesale energy, capacity and ancillary service markets. This allows distributed assets to interact with the wholesale market through the feeder's retail market and area level management systems. Price and availability information must flow from the device level to the feeder level retail aggregators and markets. The feeder level markets combine individual device bids, including battery demand/supply bids for charging/discharging, as shown in \reffig{retail_market}. 

Similarly, aggregators combine the supply bids from distributed renewables to form feeder level supply curves. The aggregate net constrained results of the demand and supply are bid to the area level management systems, which combine various feeder level bids in a similar manner to the ISO/RTO's wholesale market. Once the wholesale market clears, the cleared prices and quantities are reported back to the area and feeder level markets, which apply their local constraints through appropriate bids to clear their respective markets. 

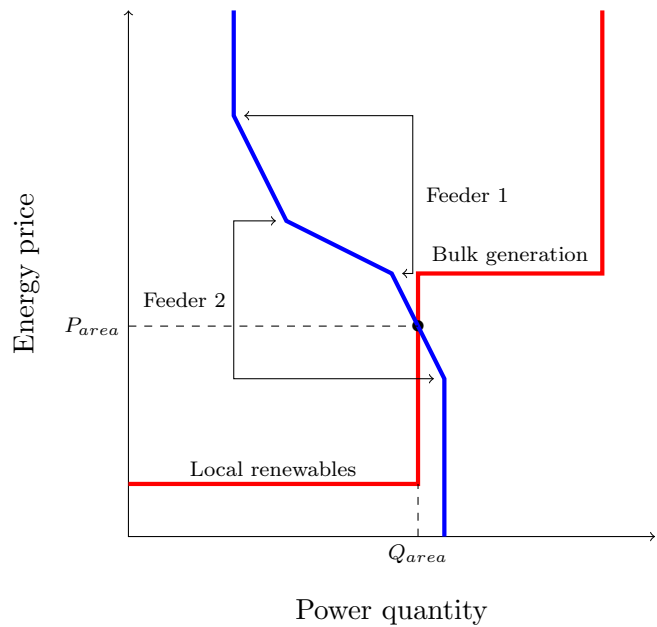
\begin{figure}[!t]
	\centering
	\begin{tikzpicture}[scale=7]
		\draw [<->] (0,1)--(0,0)--(1,0);
		\node [above,rotate=90] at (-0.15,0.5) {Energy price};
		\node [below] at (0.5,-0.1) {Power quantity};
		\draw [ultra thick,red] (0,0.1)--(0.55,0.1)--(0.55,0.5)--(0.9,0.5)--(0.9,1.0);
		\node at (0.55,0.4) {$\bullet$};
		\scriptsize
		\node [above] at (0.275,0.1) {Local renewables};
		\node [above] at (0.725,0.5) {Bulk generation};
		\draw [ultra thick,blue] (0.2,1.0)--(0.2,0.8)--(0.3,0.6)--(0.5,0.5)--(0.6,0.3)--(0.6,0.0);
		\draw [<->,black] (0.22,0.8)--(0.54,0.8)--(0.54,0.5)--(0.52,0.5);
		\node [right] at (0.55,0.65) {Feeder 1};
		\draw [<->,black] (0.28,0.6)--(0.2,0.6)--(0.2,0.3)--(0.58,0.3);
		\node [left] at (0.2,0.45) {Feeder 2};
		\draw [dashed,black] (0,0.4)--(0.55,0.4);
		\node [left] at (0,0.4) {$P_{area}$};
		\draw [dashed,black] (0.55,0)--(0.55,0.1);
		\node [below] at (0.55,0) {$Q_{area}$};
	\end{tikzpicture}
	\caption{Aggregate resource allocation using area markets. Feeders supply and demand curves are combined to determine resource participation factors.}
	\label{fig:area_market}
\end{figure}

This successive multi-level clearing process is illustrated in \reffig{area_market} by the feeder level to area level market integration. 
Based on the day-ahead wholesale market LMP, day-ahead feeder bids and other regional resources, the area market determines an area day-ahead price and available capacity of area power and establishes the area's scheduled net load and generation commitment. Thus the area forward market incorporates day-ahead supply bids from renewables and load bids from feeder to determine the day-ahead area price.  The real-time feeder markets then clear the local supply and demand to determine the feeder price that meets the area's scheduled loads, storage and distributed generation.  Storage devices then charge or discharge depending on the cleared price. In many cases end-users' bidding processes may include learning capabilities \cite{roop2005price} and must simultaneously report supply bids to participate in ancillary services markets to realize incentive compatibility under the two-part operation of those markets.

Wholesale markets must be operated by an ISO/RTO such that they facilitate the interaction between the ISO/RTO operators and the area level markets. Conventional and grid-level large-scale renewable generation fleets also operate directly in the wholesale markets. The ISO clears the wholesale market using the conventional methods of security-constrained unit-commitment and economic dispatch (SCUC/ED). 

At the area level, agents gather multiple feeder-level resources bids, combine bids and determine possible area-level resource constraints, and derive area market prices. At this level, distributed asset constraints are no longer considered, because they are embedded in the bids from the feeder level markets. In response, area controllers produce market prices that are received by the area level management system. The optimal bidding strategies problem of the area markets must be modeled as a mathematical program with equilibrium constraints. The outer problem of this bi-level problem is the area level management system's optimization problem while the inner problem is the ISO's optimization problem. The wholesale market clearing process depends on bids provided by agents and entities at all levels, but interacts only with area markets at the next level down. The bids in turn must be formulated based on market clearing processes at both wholesale and retail layers. The integration of retail and wholesale markets in this manner enables participation of distributed assets. The design of retail and wholesale markets is the same in both day-ahead and real-time markets with only the direction of information flow differing in the two markets.

\subsubsection{Inter-Temporal Integration}

Day-ahead markets are operated as pure financial markets, allowing participants to enter financially binding contracts that hedge against the price volatility of real-time markets. Because a significant majority of the energy, capacity and ancillary service resources are committed under these financial agreements, the real-time markets effectively serve as energy, capacity, and ramping imbalance markets by determining the price at which the residual must be acquired. As long as the cost of uncertainty is sufficiently higher than any emergent arbitrage opportunity, the incentive to commit a majority of resources in forward markets will be sustained and the quantities emanating from those markets will serve as reliable forecasts of supply and demand for all required resources and capabilities.  The principle of the two-settlement system will be preserved by this design and in fact could be extended for longer-term horizons if desired. 

Distributed assets must be induced to enter into contracts to procure or sell most of their energy, capacity and ramping resources in the day-ahead markets. The residual resources and needs are then submitted in the real-time markets when more precise information on the prevailing conditions becomes available. The real-time markets will also serve to correct the unforeseen imbalances between contracted day-ahead conditions and actual real-time positions. The horizontal information flow between retail and wholesale market, and their respective entities; viz., feeder-level and area-level management systems, must be addressed in a manner similar to distributed assets. 

In the presence of high levels of renewable and distributed assets, economics cannot be the only objective for utilizing distributed resources in an effective manner unless reliability can be directly translated into costs, which seems unlikely at this time. Markets and bid/control strategies for balancing services must run in parallel with retail and wholesale energy and capacity markets in both day-ahead and real-time markets. The market clearing mechanism used in the wholesale markets clear balancing reserves by co-optimizing energy and balancing needs. The balancing reserves must be determined endogenously, based on energy demand and supply, rather than set as hard limits, as is the norm in most interconnection operations today. 

\subsubsection{Balancing Services} 

The effectiveness of autonomous, grid-friendly response by smart appliances in the form of under-frequency load shedding has been already demonstrated in the Olympic project. Fifty pre-existing residential electric water heaters were retrofitted and 150 new residential clothes dryers were deployed. These responded to ambient signals received from under-frequency, load-shedding appliance controllers. Each controller monitored the frequency of the power grid's ubiquitous 60 Hz AC voltage signal at the outlet. The controllers reduced the electric load of appliance whenever electric power-grid frequency fell below 59.95 Hz. The controllers and their appliances were installed and monitored for more than a year at residential sites at three locations in Washington and Oregon. The controllers and their appliances responded reliably to each shallow under-frequency event, which occurred on average once per day and shed their loads for the durations of these events, typically about a minute and never more than ten minutes. Appliance owners reported that the appliance responses were unnoticed and caused little or no inconvenience.

There are a few more recent demonstrations of the provision of ancillary services with aggregated distributed assets such as demand response in buildings. Lawrence Berkeley National Laboratory teamed with PG\&E and SCE/Oak Ridge National Laboratory to demonstrate how residential air conditioning programs could be used to supply spinning reserves to CAISO in 250 and 2500 homes, respectively \cite{eto2002innovative}.  Ecofys led a demonstration for Bonneville Power Administration and utilities in the Pacific Northwest region to show how regulation could be provided for ``firming'' wind using water heaters, thermal storage furnaces, refrigerated warehouses, and commercial building HVAC \cite{nehrir2016making}.  A Steffes water heater and three electric vehicles are supplying regulation for PJM in ongoing demonstrations \cite{upadhye2012evaluating}. PNNL conducted an analysis that showed electric vehicles could supply all additional ancillary services for integration of 30\% of generation from wind in the Pacific Northwest region \cite{kintner2007impacts}. 

A reliability safety net created for the ISO/RTO operation can be very valuable. It would comprise fast-acting smart grid assets that can be armed and disarmed based on the clearing of a bulk-level ancillary service market and area balancing markets. These distributed assets should be aggregated into a grid-friendly network of actively managed autonomous devices that self-sense frequency and voltage fluctuations, and respond to broadcast set-point signals from control area regulation markets. These assets must provide the full range of today's ancillary services and more: virtual inertia, regulation, ramping, spinning reserve, and emergency curtailment capabilities. 

\subsubsection{Balancing Area Control}

Balancing area controllers must be able to arm and disarm distributed autonomous grid-friendly devices that provide balancing services, to reduce the burden on conventional generation, particularly when increasing balancing requirements are expected during periods of high level renewable variable generation output. At the area level, the goal of the balancing supervisor is to minimize the area control error (ACE) signal, which is a weighted sum of the deviations of the system frequency and the inter-area power flow. The balancing supervisor must coordinate with Automatic Generator Control (AGC) at the transmission level to provide frequency and tie line interchange support, minimizing balancing authority ACE. The balancing area control, in coordination with AGC, aims to maintain the system frequency at 60 Hz during the normal load demand fluctuation, and restore the system frequency gradually when a contingency occurs in the system. In particular, the balancing supervisor needs to maintain the inter-area power flow at the scheduled level. The inter-area power flow reference values are calculated based on differences of measured total area real power references (and reactive power required to realize the real power transfers) that were cleared by the real-time area market. The balancing control must balance the input of these two power references according to the current system operating conditions.  

\subsubsection{Feeder Control}

Feeder control is a critical component in the entire system because it mediates between the area control and individual device controls. Its role is bidirectional in the sense that (1) it must convert the balancing objective specified by the area control into supply constraints for the retail markets; and (2) it also must represent elasticity of those assets to the area control system. Essentially, the feeder controller has two objectives. The first is to minimize a feeder control error in the local market and the second is to enable coordination of various local devices to provide reactive power support for voltage regulation. 

A feeder's real power reference is based on the power reference cleared by the balance area control and the power reference from the retail capacity market. The feeder's balancing controller must reconcile these two power references according to the current system operating conditions. During normal conditions, the cleared power reference from the retail market will be given more weight. But during contingencies, the signal received from the balancing entity must be given more weight. 

Additionally, the feeder controller must account for variability and uncertainty of local distributed renewable resources such as rooftop photovoltaic panels, and local constraints such as feeder congestion. After the feeder real power reference is determined, the feeder control must dispatch optimal set-points to the autonomous devices that help maintain adequate power support. The feeder reactive power management system must collect real time feeder voltage information from all the devices involved in voltage support. Then the management system must coordinate with local device-level controllers by dispatching voltage setpoints and if necessary, suppressing the autonomous local control signals to avoid excessive voltage regulation.

\subsubsection{Integrated Autonomous and Market Operation} 

The real-time market management system supports both area and feeder level control but it must support one additional function. As part of the co-optimization problem at each level, ancillary service contracts must be entered into on the same time scale as the real-time energy market, and weighted according to real-time energy market requirements determined by NERC reliability standards. During normal operations, the control systems will take a purely economic perspective to maximize returns, by dispatching smart grid assets either towards real-time energy needs, ancillary service needs (such as frequency or voltage regulation, spinning reserve, etc.), or a combination of both. But during disrupted or stressed system conditions, weighting functions must be adjusted to focus on short-term system stability requirements rather than long-term economic objectives. During each real-time market cycle (e.g., every 5 minutes) the control system must establish contracts for real-time energy and balancing/regulation services, and dispatch resources subject to local constraints and availability provided by device bidding. This allows the smart grid assets to participate in multiple market revenue streams under a multi-objective control problem (i.e., storage devices participating in both energy markets and frequency regulation), capturing multiple revenue streams to increase profitability and long-term sustainability. At this time scale, only contracts for reservation of ancillary services are formed, while the control itself must be performed at a much faster rate using autonomous responses, and compensated at energy, capacity, or ramping market prices.

\subsubsection{Off-normal Operation}

Distributed smart grid assets must be pre-armed for instantaneous autonomous response during abnormal conditions, such as loss of communications or contingency events. This allows each asset to respond to the correct extent, and avoids the amplifying system-wide voltage, power, or frequency oscillations that might follow too many devices responding. For loads that cannot continuously adjust their power use (such as water heaters, HVACs), they must be switched on/off selectively so that the aggregate of a large number of these loads provides the required droop characteristic. Distributed control strategies must be designed to coordinate the different devices so they respond autonomously while maintaining the overall stability of the system. Approaches to perform this function have yet to be developed. 

\footnotesize

\bibliographystyle{ieeetr} 
\bibliography{references}

\end{document}